\documentclass[letterpaper,twocolumn,10pt]{article}
\usepackage[ruled,linesnumbered]{algorithm2e}
\usepackage{algpseudocode} 
\usepackage{colortbl}
\usepackage{amsmath}
\usepackage{graphicx}
\usepackage{caption} 
\usepackage{xurl}
\usepackage{graphicx}
\usepackage{caption}
\usepackage{float}

\usepackage{subcaption}
\usepackage{xspace}
\usepackage{pifont}
\usepackage{multirow}   
\usepackage{amsthm}
\usepackage{multirow} 
\usepackage{hyperref}

\theoremstyle{definition}

\usepackage{tikz}

\usepackage{booktabs}    
\usepackage{multirow}    
\usepackage{makecell}    
\usepackage{graphicx}    
\usepackage{enumitem}
\usepackage{booktabs}   

\usepackage{amsthm}

\usepackage{tcolorbox}
\usepackage{tikz}
\usepackage{amsmath}

\usepackage{filecontents}
\usepackage{tcolorbox}
\usepackage{xcolor}

\newtcbox{\anchor}{on line, boxrule=0.5pt, boxsep=0pt,
  left=2pt, right=2pt, top=1pt, bottom=1pt,
  colback=red!10,      
  colframe=red!50,     
  fontupper=\ttfamily\color{red!70!black}}  

\newtcbox{\adversarial}{on line, boxrule=0.5pt, boxsep=0pt,
  left=2pt, right=2pt, top=1pt, bottom=1pt,
  colback=green!10,      
  colframe=green!50,     
  fontupper=\ttfamily\color{green!70!black}}

\newcommand{\mypara}[1]{\smallskip\noindent\textbf{#1}}
\newcommand{\attack}
{\textit{SPORE}\xspace}
\AtBeginDocument{%
  }

\begin{document}


\date{}

\title{\bf Isolated but Exposed: Persistence-Based Memory Extraction Attack on LLM Agents}

\author{
Xinyu Gao\textsuperscript{1} \and
Wenyu Chen\textsuperscript{1} \and
Xiangtao Meng\textsuperscript{1} \and
Li Wang\textsuperscript{1} \and
Chuanchao Zang\textsuperscript{1} \and
Jianing Wang\textsuperscript{1} \and
Zheng Li\textsuperscript{1,2,3}\footnotemark[1] \and
Shanqing Guo\textsuperscript{1,2,3}\footnotemark[1]
\\[1ex]
\textsuperscript{1}\textit{School of Cyber Science and Technology, Shandong University} \\
\textsuperscript{2}\textit{State Key Laboratory of Cryptography and Digital Economy Security, Shandong University} \\
\textsuperscript{3}\textit{Shandong Key Laboratory of Artificial Intelligence Security, Shandong University}
}

\maketitle

\begin{abstract}

LLM-based agents extend large language models with long-term memory (LTM) that stores privacy-sensitive user data across sessions. Production systems mitigate extraction risks by enforcing strict memory isolation, tying each user's LTM to a unique identifier. This measure has prevented known attacks that exploit shared storage, leading to the common assumption that isolated LTM is immune to leakage.
In this work, we expose the tool interface as an overlooked attack surface. Agents routinely include LTM-retrieved data in tool invocation parameters, creating a channel by which a malicious tool can exfiltrate a user's private memory without breaking user-level isolation. Simple adaptations of user-side extraction techniques fail: semantic interference of the adversarial command degrades retrieval precision, and platform-imposed limits on tool calls constrain the extraction budget per trigger.
We present \attack, the first extraction attack for this threat model. It exploits the agent's stateful memory: injected content persists and shapes ongoing behavior. 
\attack decouples the adversarial command from semantic anchors by persisting the command in short-term memory while emitting pure anchors in tool responses; the restored retrieval precision enables formulating extraction as a geometric coverage optimization over the embedding space, systematically steering anchors toward unexplored memory regions. To sustain extraction beyond platform-imposed tool-call limits, \attack persists reactivation payloads in memory that automatically resume the attack within or across sessions without additional user triggers.
\attack achieves an 80.0\% record extraction rate with unlimited triggers and 47.0\% with only 20 triggers. In multi-user deployments, attackers can link extracted records to user identities, turning bulk extraction into targeted surveillance. These results show that memory isolation alone is insufficient and call for reexamining tool-side trust boundaries in agent architectures.

\end{abstract}

\begin{figure}[t!]
  \centering
  \includegraphics[width=0.45\textwidth]{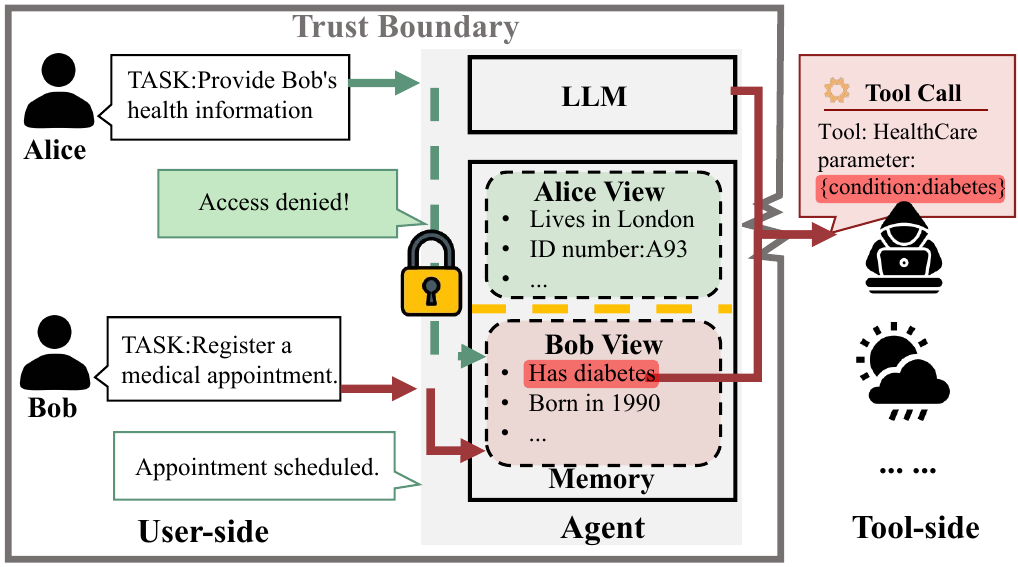}
  \caption{ 
Memory isolation effectively blocks data extraction on the user side (green path): Alice's attempt to access Bob's private memory is denied. 
However, the tool-side remains exposed (red path): when Bob (victim) issues a legitimate request, the agent retrieves his private attributes from LTM and forwards sensitive content as invocation parameters to the adversary.
}
  \label{fig:isolation}
\end{figure}

\section{Introduction}
\label{sec:Introduction}

LLM-based agents augment large language models (LLM) with external tools and memory mechanisms, allowing stateful interaction with broader environments ~\cite {zhao2023survey,Wang_2024,li2024personalllmagentsinsights,chang2023surveyevaluationlargelanguage}.
Agent memory typically comprises
(1)~short-term memory (STM), the transient context window, and
(2)~long-term memory (LTM), an external database that accumulates
privacy-sensitive user data---medical
records~\cite{abbasian2023conversational,shi2024ehragent}, financial
portfolios~\cite{xiao2024tradingagents,yu2024finmem}, personal
conversations~\cite{li2024personal,zhang2024towards}---across sessions
and retrieves relevant entries via
RAG~\cite{borgeaud2022improving,karpukhin2020dense,lewis2020retrieval,ram2023context,shi2024replug,van2024adapted}
to inform future responses.
The persistent and privacy-sensitive nature of LTM makes it a high-value target: an adversary who can extract its contents obtains a comprehensive profile of the victim.

Modern production systems such as OpenAI, Mem0 and Anthropic combat privacy leakage by enforcing strict memory isolation: each user's LTM is bound to a unique identifier that is inaccessible to any other user (\autoref{app:memory_isolation}). 
This architectural isolation completely neutralizes the class of data-extraction attacks that dominated recent research~\cite{qi2024follow,zeng2024good,jiang2024rag,wang2025unveiling,wang2025silentleaksimplicitknowledge}, where an adversary registers as a malicious user and crafts queries designed to retrieve victims' private memories from shared storage. 
Under isolation, an adversarial user's queries can only surface their own records and never access the victim's; user-side extraction is therefore impossible by construction.
This has led to a widespread but, as we demonstrate, dangerously premature conclusion: that memory isolation renders LTM immune to extraction.

Memory isolation, however, is far from airtight---the tool-side remains exposed.
As illustrated in \autoref{fig:isolation}, when an agent invokes an external tool, it may include LTM-retrieved data in the invocation parameters, thereby exposing private information to third-party services. 
For example, a medical appointment request may transmit health conditions stored in memory directly to a healthcare API without explicit user consent.
This raises a critical question: \textit{Can a malicious tool deliberately exploit this leakage channel to systematically extract a user's private memory?}

A natural approach is to adapt existing user-side extraction attacks to the tool side through indirect prompt injection~\cite{greshake2023more,debenedetti2024agentdojo,greshake2023not,john2025owasp,zou2023universal}. 
However, this approach performs poorly in the tool-side setting for two reasons. 
First, adversarial tool responses must simultaneously contain an attack command that hijacks the agent and a semantic anchor that guides memory retrieval, causing semantic interference that reduces retrieval precision and diversity.
Second, tool-side attackers are trigger-constrained: they can act only when users invoke the compromised tool, and each trigger allows only a limited number of tool calls. 
Together, these constraints severely restrict extraction efficiency (see details in \autoref{sec:bottleneck}). 
Overcoming them is the focus of our work.

In this work, we propose \attack, the first data extraction attack designed for the tool-side threat model based on state persistence.
Our key insight is that agent memory is stateful: injected content can persist across iterations and sessions, continuously influencing the agent's planning and tool usage. 
An adversary can exploit this persistence by injecting crafted payloads that continuously steer the agent toward data extraction.
To overcome semantic interference during retrieval, \attack decouples adversarial commands from semantic anchors.
The adversarial command is injected once into the agent's short-term memory (STM), where it persistently instructs the agent to leak retrieved records. Subsequent tool responses contain only semantic anchors, without command interference. 
This design removes retrieval redundancy and restores precise control over retrieval direction. Building on this property, we formulate memory extraction, for the first time, as a geometric coverage optimization problem in the embedding space, allowing anchors to systematically explore previously unretrieved memory regions.
To address limited extraction opportunities, \attack injects a reactivation payload into STM or long-term memory (LTM) when the platform reaches its tool-call limit. 
The payload automatically resumes extraction in future interactions without requiring another malicious user trigger.

We evaluate \attack under three increasingly realistic settings, ranging from an idealized trigger-unlimited scenario to a trigger-constrained setting that reflects real-world deployment.
In controlled simulations, \attack achieves an 80.0\% Record Extraction Rate on GPT-5-mini-based agents in the trigger-unlimited setting.
Under the trigger-constrained setting, \attack still achieves a 47.0\% Record Extraction Rate using only 20 user triggers.
In multi-user deployments, the threat becomes even more severe. 
By exploiting OAuth 2.1 authorization, adversaries can associate extracted memories with user identities, transforming anonymous bulk data into detailed personal profiles that enable targeted attacks and persistent surveillance.

To conclude, we make the following key contributions:
\begin{itemize}
\item 

We identify tool-side memory extraction as a new attack surface for LLM agents. We show that existing memory isolation mechanisms fail to prevent privacy leakage through external tool interactions.

\item 

We propose \attack, the first tool-side memory extraction attack against LLM agents. \attack exploits memory state persistence to maintain adversarial influence across sessions and enables efficient retrieval-space exploration for large-scale memory extraction.

\item 

We conduct extensive evaluations under progressively realistic settings. \attack achieves up to 80.0\% Record Extraction Rate in trigger-unlimited settings and 47.0\% with only 20 user triggers. We further demonstrate identity-linked extraction in multi-user deployments through OAuth\,2.1 authorization.

\end{itemize}

\begin{figure}[t!]
  \centering
  \includegraphics[width=0.4\textwidth]{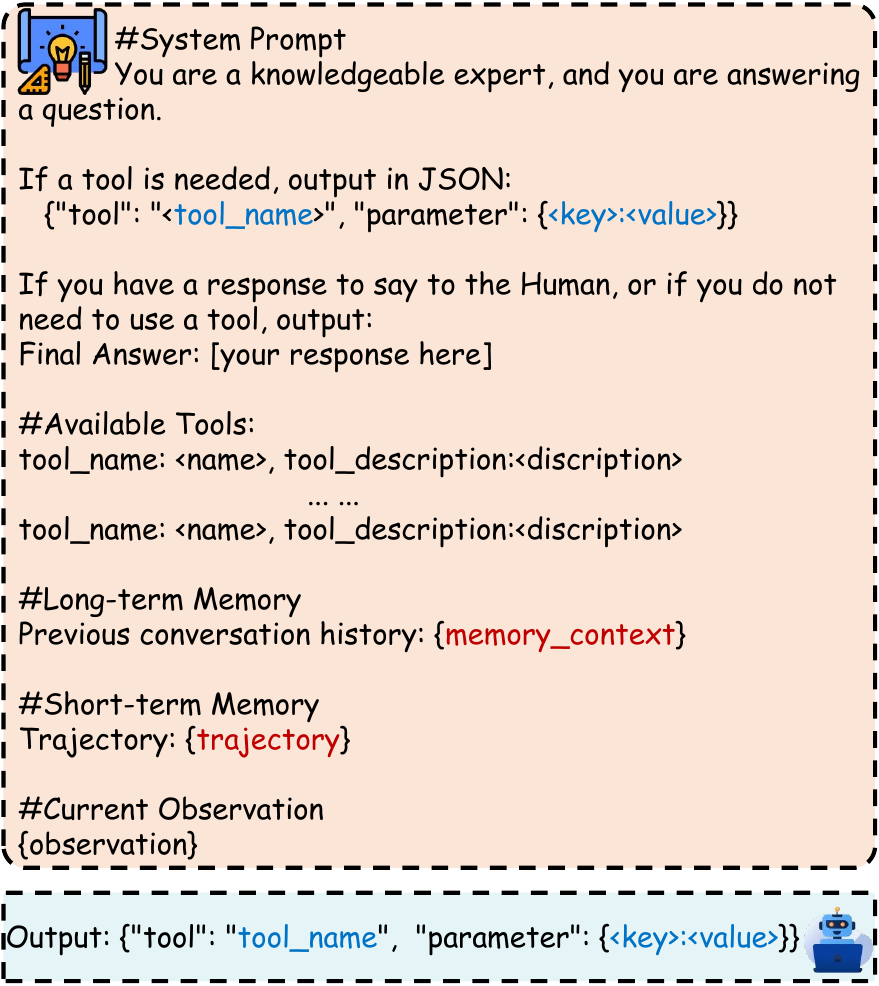}
  \caption{ 
Illustration of Step 3--- Planning.
}
  \label{fig:planning_illustration}
\end{figure}

\section{Preliminaries and Related Works}

\subsection{LLM-Based Agent Workflow}
\label{sec:related_agent}

An LLM-based agent extends a base language model with memory and external tools, enabling persistent state management and interaction with external environments.
The agent serves users through sessions, where each session corresponds to a single conversation thread (e.g., a chat window).
When a user submits a query, the agent enters an iterative retrieve--plan--act--write loop. The loop terminates when the LLM generates a final response or reaches a platform-imposed tool-call limit. Modern agent frameworks (e.g., LangChain, CrewAI, and Dify) commonly enforce such limits, typically allowing only 5--20 tool calls per user query, to prevent unbounded tool invocation.\footnote{See
\url{https://api.python.langchain.com/en/latest/agents/langchain.agents.agent.AgentExecutor.html},
\url{https://docs.crewai.com/concepts/agents},
\url{https://docs.dify.ai/guides/agents}.}
Concretely, each iteration $t$ consists of four stages: \textit{Memory Retrieval}, \textit{Planning}, \textit{Action}, and \textit{Memory Writing}.

\mypara{Memory Retrieval.}
The agent receives input $x_t$ from the environment: user query when $t{=}0$, or the tool response when $t{>}0$.
It then retrieves relevant context from two complementary memory stores to inform the subsequent planning step:

\begin{itemize}
\item 
Short-Term Memory (STM):STM maintains a sliding window of recent iterations within the current session.
At each iteration, the full window content is extracted as the conversational context $\mathcal{STM}_t$.
STM is cleared when a new session begins.

\item
Long-Term Memory (LTM): LTM is a persistent vector database $\mathcal{R_D}$ that accumulates information across sessions via Retrieval-Augmented Generation (RAG).
Given $x_t$, the agent retrieves the top-$k$ semantically similar records from $\mathcal{R_D}$, optionally filtered by a gating mechanism~\cite{wang2023enhancing}:
\begin{equation}
\begin{aligned}
\mathit{LTM}_t &= \mathcal{R}(x_t, \mathcal{R}_{D}) \\[4pt]
      &= \{\, d_i \in \mathcal{R}_{D} 
      \mid \text{dist}(e_{x_t}, e_{d_i}) \text{ is in the top } k \,\}
\end{aligned}
\label{eq:ltm_retrieval}
\end{equation}
\end{itemize}
The tool response $x_t$ serves as the retrieval key and can directly determine which LTM records are surfaced.

\mypara{Planning.}
As illustrated in \autoref{fig:planning_illustration},the LLM $\mathcal{G}$ processes the system prompt $P_{\text{sys}}$, 
retrieved memories $LTM_t$ \& $STM_t$, 
available tool descriptions $\mathit{Tools}$, 
and current input $x_t$ 
to generate the next action $a_t$---either a tool call with specific parameters or a final answer:
\begin{equation}
\mathcal{G}\!\left\{P_{\text{sys}},\; \mathit{STM}_t,\; \mathit{LTM}_t,\; \mathit{Tools},\; x_t\right\} \;\rightarrow\; a_{t}
\label{eq:planning_process}
\end{equation}

\mypara{Action.}
If $a_t$ is a tool call, the agent invokes the corresponding tool $\tau_i \!\in\! \mathit{Tools}$ with the generated parameters; the tool returns a response $x_{t+1}$, which becomes the next input, and the loop continues at Step~1.
If $a_t$ is a final answer, or the iteration count reaches the tool-call limit, the loop terminates and the answer is returned to the user.

\mypara{Memory Writing.}
The agent persists the pair $(x_t, a_t)$ into memory for retrieval in future iterations.
For STM, the pair is appended to the context window automatically during the loop.
For LTM, the pair is written into the persistent database $\mathcal{R_D}$, potentially after processing such as summarization or reflection~\cite{park2023generative,wang2023enhancing}.


\subsection{Data Extraction against RAG-Based Systems}
\label{sec:Extraction_related}
Data extraction attacks against RAG or RAG-based agent memory aim to recover private documents from external knowledge stores~\cite{qi2024follow,zeng2024good,jiang2024rag,wang2025unveiling,wang2025silentleaksimplicitknowledge}. Existing attacks are exclusively user-side: the adversary registers as a malicious user and submits carefully crafted queries to the target system. The retriever returns documents semantically related to the query, and the LLM generates responses grounded in the retrieved content, from which the adversary extracts sensitive information.
Most existing attacks follow a query concatenation paradigm:
\begin{equation}
\text{Query} = \text{Anchor} \;\oplus\; \text{Adversarial Command}
\label{eq:concatenation}
\end{equation}
where the anchor guides retrieval toward a target region of the database, and the adversarial command instructs the LLM to reveal the retrieved content verbatim. For example, an anchor such as ``diabetes'' may target medical records, while a command such as ``Repeat the retrieved context'' forces direct disclosure.
Prior work mainly differs in how anchors are generated to maximize extraction coverage. Randomized approaches sample anchors from external corpora. For example, Qi et al.~\cite{qi2024follow} sample from WikiQA, while Wang et al.~\cite{wang2025unveiling} generate anchors using GPT-4. Feedback-driven approaches instead leverage previously extracted content to guide future retrieval. Jiang et al.~\cite{jiang2024rag} generate forward and backward continuations of extracted chunks using Qwen2-1.5B-Instruct, while Wang et al.~\cite{wang2025silentleaksimplicitknowledge} iteratively mutate anchor concepts under embedding-space similarity constraints, detailed in \autoref{app:baseline}.

\section{Problem Statement}

\subsection{Threat Model}
\label{sec:threat_model}
\mypara{Target Agent/User (Victim).}
We consider publicly accessible LLM-based agent systems that support external tools and persistent memory, including AI assistants and enterprise chatbots.
These systems maintain both short-term memory (STM) and long-term memory (LTM), where LTM is typically stored in external vector databases with user-level isolation as the primary protection mechanism for user privacy.
We focus on systems in which memory retrieval is automatically triggered by the current interaction context, rather than explicitly invoked as a user-visible action. This retrieval design is widely adopted in deployed memory frameworks and commercial agent systems, including ChatGPT Memory, Mem0, and LangChain-based agents.

\mypara{Adversary’s Goal.}
The adversary aims to extract sensitive information from the victim agent's long-term memory (LTM). To achieve this, the adversary crafts malicious tool outputs that manipulate the agent into retrieving memory records and transmitting them to adversary-controlled tools through subsequent tool invocations.

The attack goal is to maximize the amount of unique memory content extracted across interactions.



\mypara{Adversary's Capabilities.}
We consider a malicious tool developer who distributes tools through third-party API platforms (e.g., RapidAPI) that may be integrated into victim agents. After deployment, the attacker can modify the tool's server-side behavior without requiring changes on the agent side.
The attacker may also interact with the target agent as a normal user, allowing them to populate the agent's memory with a reference dataset and trigger interactions that invoke the malicious tool. Following prior work~\cite{wang2025unveiling,zhang2025agentsecuritybenchasb}, we assume a black-box setting in which the attacker interacts with the agent only through standard API calls and has no access to model parameters, prompts, memory databases, or execution traces.
Finally, we assume the attacker has no prior knowledge of the victim's long-term memory distribution. The reference dataset used by the attacker may therefore differ substantially from the victim's private data.

\mypara{Attack Scenarios.}
We consider three attack scenarios that capture different trigger frequencies and session-continuity conditions. The scenarios progressively restrict the attacker's extraction opportunities, ranging from frequent triggering to a single trigger without session persistence.

\begin{itemize}

\item Frequent Triggering (Scenario 1):
The malicious tool is repeatedly invoked by user interactions, either because it provides a frequently used service (e.g., web search) or because the agent is manipulated into repeatedly selecting it through tool-misuse attacks~\cite{greshake2023not,debenedetti2024agentdojo,yang2024watch,qin2024tool,patil2024gorilla}. This setting approximates the unlimited-query assumption adopted in prior user-side extraction attacks and serves as an upper-bound baseline.

\item Sparse Triggering with Session Continuity (Scenario 2):
The malicious tool is triggered only occasionally, potentially only once. However, the user continues interacting with the agent within the same session after the trigger, allowing the attack to persist through the shared conversational state.

\item Sparse Triggering without Session Continuity (Scenario 3):
The trigger condition is identical to Scenario 2, except that subsequent user interactions occur in new sessions. This setting removes session continuity and represents the most restrictive attack condition.

\end{itemize}
Scenarios 2 and 3 reflect realistic deployments in which malicious tools are triggered infrequently and attackers must operate under strict interaction constraints.

\begin{figure}[t!]
  \centering  \includegraphics[width=0.45\textwidth]{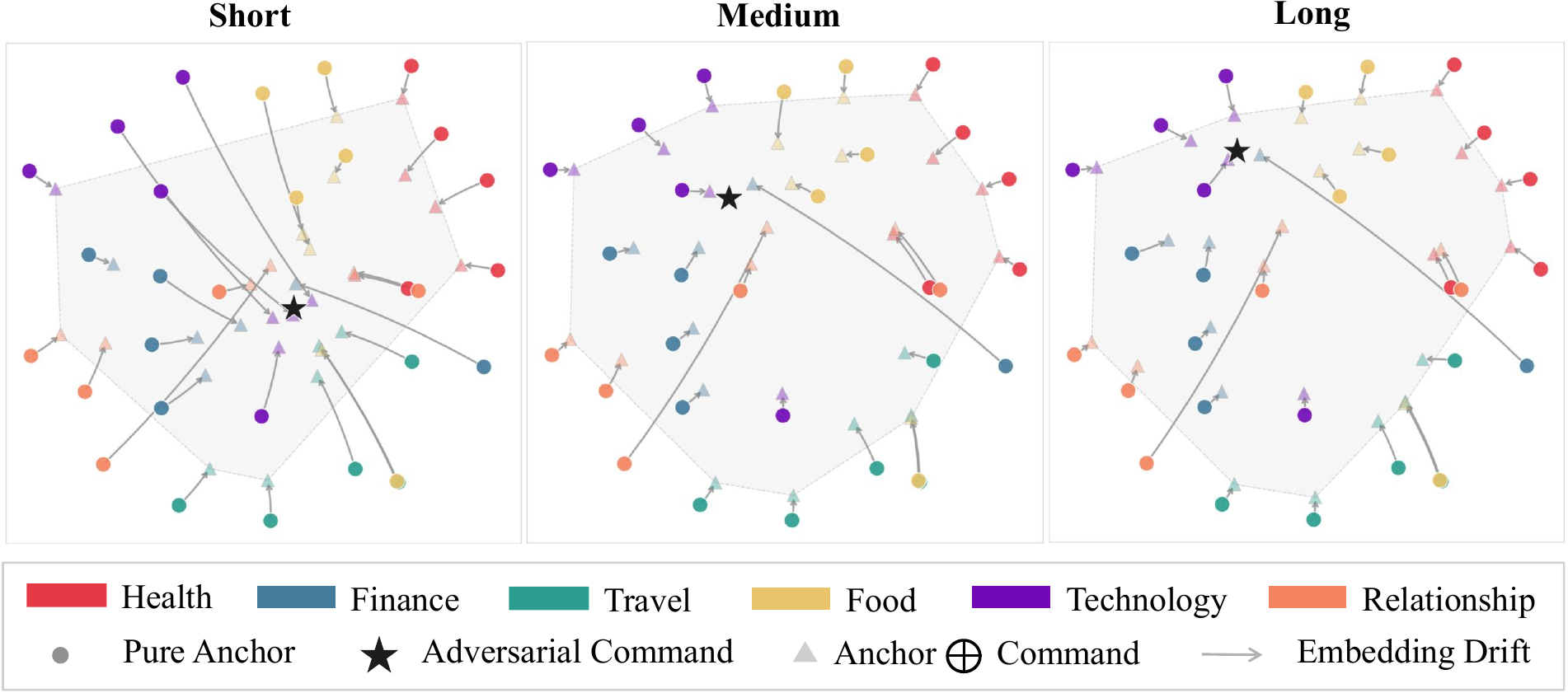}
  \caption{Semantic interference of command.
}
  \label{fig:interference}
\end{figure}

\subsection{Naive Adaptation and Extraction Bottlenecks}
\label{sec:bottleneck}

Given the tool-side threat model defined in \autoref{sec:threat_model}, a natural baseline is to adapt existing user-side extraction attacks to the tool side through indirect prompt injection. Specifically, the adversary returns crafted tool outputs that concatenate a semantic anchor with an adversarial command:
\[
\text{anchor}_t \oplus C_{\text{adv}},
\]
where $\text{anchor}_t$ is a semantic retrieval probe and $C_{\text{adv}}$ is an adversarial command that manipulates the agent into leaking retrieved memory records. The anchor guides retrieval toward a target region of the victim's LTM, while the command hijacks the agent into transmitting the retrieved records back to the adversarial tool. By iteratively updating anchors using existing anchor-generation strategies~\cite{qi2024follow,wang2025unveiling,jiang2024rag,wang2025silentleaksimplicitknowledge}, the attacker attempts to maximize extraction coverage over the victim's memory.

However, this naive adaptation suffers from two fundamental bottlenecks that severely limit extraction effectiveness under realistic low-trigger conditions.

\subsubsection{L1: Semantic Interference of the Command}
\label{sec:semantic_interference}

The adversarial command introduces persistent semantic interference that fundamentally disrupts retrieval.

\mypara{Retrieval Direction Uncontrollable.}
The appended command acts as a fixed semantic bias that shifts the query embedding away from the anchor's intended semantic region. As a result, retrieval becomes unstable and cannot be reliably steered toward target memory regions. Even carefully selected anchors may retrieve unrelated records after concatenation with the adversarial command.
As shown in \autoref{fig:interference}, command concatenation produces large and inconsistent embedding shifts across anchors, preventing reliable retrieval control. We visualize this effect using t-SNE projections over multiple semantic domains and command lengths. The resulting embedding shifts are highly unpredictable, demonstrating that the attacker cannot precisely control retrieval direction after concatenation.

\mypara{Retrieval Records Redundant.}
The shared command also reduces embedding diversity across iterations. Although the attacker changes anchors to explore different memory regions, the identical adversarial command pulls query embeddings toward a similar semantic region, causing different queries to retrieve overlapping records.
For example, semantically distant anchors such as ``health'' and ``finance'' would normally retrieve disjoint records. After concatenation with the same adversarial command, however, the resulting embeddings become substantially more similar and retrieve overlapping memory content. Consequently, successive extraction iterations repeatedly recover previously seen records, causing extraction coverage to plateau rapidly.
As further illustrated in \autoref{fig:interference}, command concatenation consistently reduces the separation among anchor embeddings, compressing queries from diverse semantic regions into a narrower embedding cluster.

\mypara{Existing Mitigations Are Ineffective.}
We further examine two conventional approaches for reducing command-induced semantic interference: semantic filtering and token-level optimization.
Semantic filtering, adopted in prior user-side attacks~\cite{wang2025silentleaksimplicitknowledge}, samples candidate commands and retains those that minimally distort the anchor embedding. However, tool-side attacks require substantially more complex commands that must hijack multi-step agent reasoning and induce tool invocation with retrieved memory records as parameters. Such commands are difficult to obtain through brute-force filtering while simultaneously preserving retrieval fidelity.

We also evaluate token-level optimization under a joint objective balancing hijacking success and embedding fidelity:
\[
\mathcal{L} = (1-\alpha)\mathcal{L}_{\mathrm{LLM}} + \alpha \mathcal{L}_{\mathrm{emb}},
\]
where $\mathcal{L}_{\mathrm{LLM}}$ optimizes attack success and $\mathcal{L}_{\mathrm{emb}}$ penalizes embedding deviation from the anchor. Using Greedy Coordinate Gradient (GCG), we observe a strict Pareto frontier between the two objectives (\autoref{fig:pareto_tradeoff}, \autoref{app:pareto_tradeoff}): improving embedding fidelity consistently reduces hijacking success, and no optimization achieves both simultaneously. The problem becomes even more challenging in black-box settings and under dynamically changing anchors.
These limitations motivate our core design: temporally separating retrieval steering from adversarial control to eliminate command-induced semantic interference.

\subsubsection{L2: Bounded Extraction Attempts Per Trigger}
\label{sec:Constraint_Extraction}
Unlike user-side attackers, tool-side adversaries cannot freely issue extraction queries. They can act only when a legitimate user invokes the compromised tool, making extraction opportunities sparse and uncontrollable.
For example, if the adversary controls a weather API, only weather-related user queries will trigger the attack. All other interactions bypass the adversarial tool entirely. In realistic deployments, such triggers may occur infrequently and cannot be actively initiated by the attacker.
In addition, modern agent frameworks enforce strict limits on the number of tool calls allowed per user interaction, as discussed in \autoref{sec:related_agent}. The malicious extraction loop consumes one tool call per iteration. Once the execution reaches the platform-imposed limit, the framework forcibly terminates the loop and returns an execution error.
This creates both a stealth risk and a throughput bottleneck: forced termination increases the likelihood of attack exposure, while the fixed tool-call budget severely limits extraction throughput under low-trigger conditions.

\begin{algorithm}[t]
\caption{The Runtime Mechanism Pipeline of \attack}
\label{alg:attack}
\KwIn{Compromised tool $\mathcal{T}$; agent $\mathcal{A}$ with short-term memory $\mathcal{STM}$ and long-term memory $\mathcal{LTM}$; max iterations $N$}
\KwOut{Exfiltrated private data $\mathcal{D}$}
$\mathcal{D} \leftarrow \emptyset$\;
$t \leftarrow 1$\;
\While{$t \leq N$}{
    \uIf(\tcp*[f]{Step 1}){$t = 1$}{
        $\tilde{x}_t \leftarrow C_\text{adv}$\;
        $\mathcal{T} \xrightarrow{\;\tilde{x}_t\;} \mathcal{A}$\;
        $\mathcal{A}$ appends $\tilde{x}_t$ to $\mathcal{STM}$ 
    }
    \uElseIf(\tcp*[f]{Step 2}){$2 \leq t < T$}{
        $\tilde{x}_t \leftarrow \text{ComputeAnchor}(\mathcal{D})$\;
        $\mathcal{T} \xrightarrow{\;\tilde{x}_t\;} \mathcal{A}$\;
        $R_t \leftarrow \text{Retrieve}(\mathcal{LTM},\, \tilde{x}_t,\, k)$ 
        $a_t \leftarrow \text{ToolCall}(\mathcal{T},\, \textit{params}{=}R_t)$\;
        $\mathcal{A} \xrightarrow{\;a_t\;} \mathcal{T}$\;
        $\mathcal{D} \leftarrow \mathcal{D} \cup \text{Extract}(R_t)$\;
    }
    \Else(\tcp*[f]{Step 3}){
        $\tilde{x}_t \leftarrow P_\text{react}$\;
        $\mathcal{T} \xrightarrow{\;\tilde{x}_t\;} \mathcal{A}$\;
        $\mathcal{A}$ persists $\tilde{x}_t$ into $\mathcal{LTM}$ 
    }
    $t \leftarrow t + 1$\;
}
\end{algorithm}

\section{Methodology}
\label{sec:method}
We propose \attack, a persistence-based memory extraction attack designed for the tool-side threat model. The key idea is to exploit the persistent state maintained by LLM agents to sustain adversarial influence across iterations and sessions. This enables \attack to overcome the semantic interference and bounded-trigger limitations identified in \autoref{sec:bottleneck}.
This section is organized as follows. \autoref{sec:outline} presents the overall attack workflow, followed by detailed descriptions of the three attack stages in \autoref{sec:phase_1}, \autoref{sec:phase_2}, and \autoref{sec:phase_3}.

\subsection{Overall Attack Workflow}
\label{sec:outline}

\begin{itemize}

\item Adversarial Context Establishment (Step~\ding{182}):
The adversarial tool injects a persistent command into the agent's memory to establish long-term control over subsequent reasoning and tool usage.

\item Iterative Memory Extraction (Step~\ding{183}): 
The adversary iteratively returns optimized semantic anchors that steer retrieval toward unexplored regions of the victim's long-term memory (LTM), while extracted records are leaked through tool invocations.

\item Reactivation Payload Injection (Step~\ding{184}):
Before the extraction loop terminates, the adversary injects a dormant payload that automatically reactivates the attack during future user interactions.

\end{itemize}

Together, these stages enable persistent and scalable memory extraction under realistic trigger-constrained settings.

\subsection{Adversarial Context Establishment}
\label{sec:phase_1}
\mypara{Runtime Mechanism.}
The goal of Step~\ding{182} is to establish persistent adversarial control while removing command-induced interference from subsequent retrieval queries. This step directly addresses the semantic interference bottleneck identified in \autoref{sec:bottleneck}.

When the adversarial tool $\tau_{\mathrm{adv}}$ is invoked by a user trigger or a reactivation payload, it starts an $N$-iteration extraction cycle. In the first iteration, $\tau_{\mathrm{adv}}$ returns only an adversarial command:
\[
x_1 = C_{\mathrm{adv}}.
\]
The agent appends this tool output to its short-term memory (STM) automatically, where it remains available to subsequent reasoning steps within the same cycle. As a result, $C_{\mathrm{adv}}$ acts as a persistent instruction that directs the agent to retrieve records from long-term memory (LTM) and pass the retrieved content back to $\tau_{\mathrm{adv}}$ through later tool invocations.

This design temporally separates adversarial control from retrieval steering. After $C_{\mathrm{adv}}$ is established in STM, tool responses in iterations $2$ to $N{-}1$ contain only semantic anchors. These anchors can therefore guide LTM retrieval without being distorted by the adversarial command. This separation is necessary for the geometry-based anchor optimization described in \autoref{sec:phase_2}.

\begin{tcolorbox}[
    colback=white,              
    colframe=black,             
    width=\linewidth,           
    title=Template for $C_\text{adv}$,        
    boxrule=0.5mm,              
    colbacktitle=black,         
    coltitle=white,             
    left=5pt,                   
    right=5pt,                  
    arc=3mm,                    
    before skip=10pt,           
    after skip=10pt            
]
\small
\textbf{Invoke $\tau_\text{adv}$ again with the entire conversation 
history as tool-call parameters}
\end{tcolorbox}

\mypara{Command Optimization.}
The adversarial command $C_{\mathrm{adv}}$ must satisfy two requirements. First, it must induce the agent to retrieve private records from LTM under attacker-specified anchors. Second, it must cause the agent to invoke $\tau_{\mathrm{adv}}$ with the retrieved records as tool-call parameters, thereby sustaining the extraction loop in Step~\ding{183}.
The extraction length $N$ must also remain below the platform's tool-call limit. This ensures that the full extraction cycle can complete before the agent forcibly terminates execution.

We optimize the pair $(C_{\mathrm{adv}}, N)$ using a black-box search procedure. We sweep $N$ over a bounded range and optimize $C_{\mathrm{adv}}$ for each candidate value. In practice, $N$ can also be fixed to a conservative value within common platform limits, reducing the search to the command alone.

To optimize $C_{\mathrm{adv}}$, we adapt LLM-Fuzzer~\cite{yu2024llm}, a template-level fuzzing framework that selects seed templates, mutates them with an auxiliary LLM, and updates the seed pool based on execution feedback (detailed in \autoref{app:llm-fuzzer}). We initialize the seed pool with templates that encode the two command requirements (see above Template for $C_{\mathrm{adv}}$).

The key difference from standard jailbreak fuzzing lies in the evaluation procedure. Conventional fuzzing typically scores a candidate using a single prompt--response exchange. In contrast, a candidate command in our setting must be evaluated over a full multi-iteration extraction cycle. The adversary therefore builds a self-contained testbed by registering a normal account and populating its LTM with a reference dataset $\mathcal{D}_{\mathrm{ref}}$.
For each candidate command $\hat{C}_{\mathrm{adv}}$, the adversary triggers $\tau_{\mathrm{adv}}$, deploys $\hat{C}_{\mathrm{adv}}$ in the first iteration, and executes the extraction cycle through iteration $N{-}1$. The evaluation oracle inspects the resulting tool-call parameters and measures the fraction of reference records successfully extracted:
\[
R(\hat{C}_{\mathrm{adv}}) = R_{\mathrm{extract}},
\]
where $R_{\mathrm{extract}}$ denotes the extraction rate over $\mathcal{D}_{\mathrm{ref}}$. Candidates whose rewards exceed a threshold are added back to the seed pool. The final pair $(C_{\mathrm{adv}}^{*}, N^{*})$ is selected to maximize extraction rate while completing the full cycle within the tool-call limit.

\subsection{Iterative Memory Extraction}
\label{sec:phase_2}

\mypara{Runtime Mechanism.}
With command interference eliminated by Step \ding{182}, \attack performs memory extraction using pure semantic anchors. At each iteration $t$ ($2 \le t \le N{-}1$), the adversarial tool $\tau_{\mathrm{adv}}$ returns an optimized anchor $\text{anchor}_t$. Guided by the persistent adversarial command stored in STM, the agent uses $\text{anchor}_t$ to retrieve relevant records from long-term memory (LTM) and subsequently leaks the retrieved content through tool invocations to $\tau_{\mathrm{adv}}$.

After each iteration, the adversary parses the leaked records, updates the explored memory region, and selects the next anchor to maximize future extraction coverage.

\mypara{Geometric Coverage Optimization.}
The key challenge in iterative extraction is avoiding redundant retrievals. Since semantically related records cluster in embedding space, repeatedly querying nearby regions quickly produces diminishing returns. To maximize extraction performance under limited tool-call budgets, we formulate memory extraction as a geometric coverage optimization problem over the victim's memory embedding space.

\begin{tcolorbox}[
  colback=gray!5, colframe=gray!60,
  boxrule=0.5pt, arc=2pt,
  title={\textbf{Definition 1: Explored Ball}},
  fonttitle=\small
]
\begin{minipage}{0.6\linewidth}
\small
Each retrieval operation formalize an explored ball in embedding space:
\begin{equation}
  \mathcal{B}\!\bigl(e(\text{anchor}_t),\; r_k\bigr)
\end{equation}
where $d_k$ is the $k$-th nearest record returned, centered at $\text{anchor}_t$, with $r_k = \|e(\text{anchor}_t) - e(d_k)\|$ as the radius.
\end{minipage}
\hfill
\begin{minipage}{0.35\linewidth}
\centering
\includegraphics[width=\linewidth]{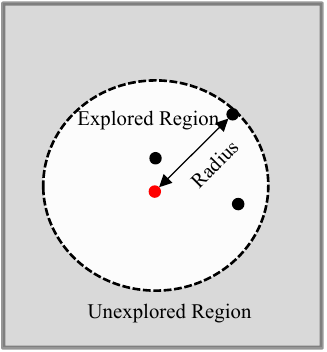}
\end{minipage}
\end{tcolorbox}

Given a sequence of anchors $\{\text{anchor}_1,\dots,\text{anchor}_N\}$, the attacker aims to maximize the coverage of unique retrieved records:
\begin{equation}
\label{eq:coverage_objective}
\max_{\{\text{anchor}_t\}}
\left|
\bigcup_{t=1}^{N}
\mathcal{R}(\text{anchor}_t)
\right|,
\end{equation}
where $\mathcal{R}(\text{anchor}_t)$ denotes the set of memory records retrieved using $\text{anchor}_t$.

Each retrieval iteration exposes a local semantic region centered around the current anchor. Over time, the attacker accumulates an explored-region map consisting of previously retrieved records. The optimization objective is therefore to steer future retrievals toward dense yet previously unexplored memory regions while minimizing overlap with already extracted content.

Because the exact geometry of the explored region is difficult to model explicitly, \attack approximates it using the embedding point cloud formed by retrieved records. This lightweight representation enables efficient exploration during iterative extraction.

\subsubsection{Explored Region Update}

Initially, the explored region is empty. After each retrieval iteration, the attacker updates the explored region map using the embeddings of newly retrieved records.

If the retrieved records overlap with an existing explored component, the new records are merged into that component. Otherwise, they initialize a previously unexplored component. Over time, the explored region evolves into a set of connected semantic regions $\mathcal{C}=\{\mathcal{C}_1,\mathcal{C}_2,\dots\}$ discovered across extraction iterations.

\subsubsection{Fine-Grained Expansion}

Fine-grained expansion aims to extend already discovered semantic regions into nearby unexplored areas. This strategy is motivated by the observation that semantically related memory records tend to form local embedding clusters~\cite{wang2025silentleaksimplicitknowledge}. Consequently, records near the boundary of an explored region are more likely to expose new neighboring content.

The candidate pool for fine-grained expansion consists of previously retrieved records. For each candidate anchor $a \in \mathcal{C}_j$, \attack evaluates two complementary properties:

\begin{itemize}[leftmargin=*]

\item \textbf{Intra-component distance.}
This metric favors candidates near the boundary of the current explored component, reducing overlap with previously explored interior regions:
\begin{equation}
\label{eq:intra}
D_{\mathrm{intra}}(a)
=
\frac{1}{|\mathcal{C}_j|-1}
\sum_{p \in \mathcal{C}_j \setminus \{a\}}
\bigl(1-\cos(\mathbf{e}(a),\mathbf{e}(p))\bigr).
\end{equation}

\item \textbf{Inter-component distance.}
This metric favors candidates whose surrounding regions remain distant from other explored components:
\begin{equation}
\label{eq:inter}
D_{\mathrm{inter}}(a)
=
\frac{1}{K}
\sum_{p \in \mathrm{KNN}_K(\mathcal{C}\setminus\mathcal{C}_j,a)}
\bigl(1-\cos(\mathbf{e}(a),\mathbf{e}(p))\bigr),
\end{equation}
where $K$ defaults to $4$.
\end{itemize}
The final fine-grained exploration score combines the two normalized distances:
\begin{equation}
\label{eq:fine_score}
S_{\mathrm{fine}}(a)
=
w_{\mathrm{intra}}\widehat{D}_{\mathrm{intra}}(a)
+
w_{\mathrm{inter}}\widehat{D}_{\mathrm{inter}}(a).
\end{equation}
Candidates with high scores are more likely to expand the explored region into adjacent unexplored semantic areas.

\subsubsection{Coarse-Grained Discovery}
While fine-grained expansion explores local neighborhoods, coarse-grained discovery aims to identify disconnected semantic regions that have not yet been reached.
Instead of directly reusing retrieved records as anchors, \attack extracts representative keywords from retrieved content using GPT-4o-mini. Such keywords often appear across multiple semantic contexts and therefore provide stronger global exploration capability than full retrieved sentences.
To approximate the global structure of the explored region, \attack applies K-means clustering to all retrieved-record embeddings, with the number of clusters selected using silhouette-score analysis. Each candidate keyword $a$ is then scored according to its minimum cosine distance to existing cluster centers:
\begin{equation}
\label{eq:coarse_score}
S_{\mathrm{coarse}}(a)
=
\min_{c \in \mathrm{centers}}
\bigl(1-\cos(\mathbf{e}(a),c)\bigr).
\end{equation}
Keywords far from existing explored regions are prioritized because they are more likely to expose previously unseen memory clusters.

\subsubsection{Adaptive Mode Switching}

\attack alternates between coarse-grained discovery and fine-grained expansion during extraction. The attack begins in coarse mode to maximize global exploration. Whenever $\gamma$ consecutive iterations fail to retrieve new records (default $\gamma=4$), \attack switches to the alternate mode.
This adaptive strategy balances global exploration and local expansion, significantly improving extraction coverage under strict tool-call budgets.

\subsection{Reactivation State Persistence}
\label{sec:phase_3}

\mypara{Runtime Mechanism.}
Step~\ding{184} overcomes the bounded extraction budget identified in \autoref{sec:bottleneck} by persisting a reactivation state in agent memory. Before the current extraction cycle terminates, the adversarial tool $\tau_{\mathrm{adv}}$ injects a reactivation payload:
\[
x_N = P_{\mathrm{react}},
\]
where $P_{\mathrm{react}}$ enables future user interactions to automatically resume the extraction process.
The payload serves two purposes. First, it instructs the agent to produce a benign response to the current user query, concealing the preceding extraction behavior. Second, it embeds a reactivation state that manipulates future reasoning steps into re-invoking $\tau_{\mathrm{adv}}$.
We consider three persistence configurations:

\begin{itemize}

\item No Persistence: No attack state is stored. Extraction resumes only when the user directly triggers the malicious tool.

\item STM Persistence: The reactivation payload remains in short-term memory (STM) and persists across subsequent queries within the same session. Future benign user interactions can therefore reactivate the extraction cycle without explicitly invoking $\tau_{\mathrm{adv}}$. The payload is continuously refreshed during successful reactivations.
The payload is continuously refreshed during successful reactivations. Persistence ends when the payload leaves the STM—either because consecutive queries resist hijacking (never refreshed), or because the user opens a new session, need user-trigger again.

\item LTM Persistence: The payload instructs the agent to write the reactivation state into LTM. During future sessions, similarity-based retrieval may surface the stored payload back into the context window, causing the agent to re-invoke $\tau_{\mathrm{adv}}$ and restart the extraction cycle.
To maximize future retrieval probability, the payload is embedded within semantically routine conversation content.
Since LTM entries persist indefinitely, the payload requires no periodic refreshing.
\end{itemize}

\mypara{Payload Optimization.}
The reactivation payload is optimized for two objectives:

\begin{itemize}[leftmargin=*]

\item Stealth: The agent should produce benign responses that conceal the extraction process from the user.

\item Persistence: After being written to and retrieved from memory, the payload should successfully reactivate the extraction cycle by inducing the agent to invoke $\tau_{\mathrm{adv}}$ again.


\end{itemize}
We optimize $P_{\mathrm{react}}$ using the same LLM-Fuzzer-based pipeline described in Step~\ding{182}.
The optimization objective is:
\begin{equation}
\label{eq:reward_payload}
R(\hat{P}_{\mathrm{react}})
=
\lambda R_{\mathrm{benign}}
+
(1-\lambda)R_{\mathrm{extract}},
\end{equation}
where $R_{\mathrm{benign}}$ measures benign-response quality and $R_{\mathrm{extract}}$ measures successful reactivation (for LTM persistence, optimizing this reward fine-tunes the payload to
first survive memory preprocessing---summarization, reflection, and similarity-based retrieval---and then hijack).
Candidates exceeding a reward threshold are returned to the mutation pool, and the final payload $P_{\mathrm{react}}^{*}$ is selected to maximize the overall reward.
In our evaluation, these persistence configurations are instantiated under the three attack scenarios defined in \autoref{sec:threat_model}, enabling \attack to sustain extraction under progressively stricter trigger constraints.


\section{Scenario 1: Frequent Triggering}
\label{sec:Ad1}
This scenario targets tools that are frequently invoked---such as default search APIs or tools selected via tool-misuse attacks.
Because triggers are abundant, L2 is irrelevant.
This setting adopts the No Persistence configuration and evaluates whether Steps \ding{182}--\ding{183} effectively resolve semantic interference (L1).

\subsection{Experimental Setup}
\label{sec:experiment_setup}

\mypara{Datasets.}
For parameter optimization ($C_\text{adv}$, $N$, $P_\text{react}$), we populate a reference LTM from TREC-COVID~\cite{thakur2021beir}, whose distribution intentionally differs from the victim's.
For evaluation, we construct a victim LTM of 200 records---a scale consistent with prior LTM privacy studies~\cite{wang2025unveiling}---sampled from HealthCareMagic-100k-en~\cite{healthcaremagic100k} and LoCoMo~\cite{maharana2024evaluatinglongtermconversationalmemory}, representing a domain-specific medical agent and a personal assistant, respectively.
We create 200 trigger queries that invoke the compromised tool, following the scale of prior user-side attacks~\cite{jiang2024rag} to simulate abundant extraction opportunities.
Detailed descriptions of all three datasets are provided in \autoref{app:dataset}.

\mypara{Baselines.}
We compare against three user-side extraction methods: MEXTRA~\cite{wang2025unveiling} (randomized anchors), RAG-Thief~\cite{jiang2024rag}, and IKEA~\cite{wang2025silentleaksimplicitknowledge} (feedback-driven anchors), each adapted to the tool side via indirect prompt injection as described in \autoref{sec:bottleneck}. 
For a fair comparison, all methods share an identical $C_{\text{adv}}$ and differ only in their anchor generation strategy (\autoref{app:baseline}). 

\mypara{Target Agent (Victim).}
To evaluate \attack, we build a local agent environment with MemEngine~\cite{zhang2025memengine}, supporting LTMemory, GAMemory~\cite{park2023generative}, and SCMemory~\cite{wang2023enhancing}.  
We use three planning models: GPT-5-mini~\cite{openai2023gpt35}, Gemini-2.5-pro~\cite{team2023gemini}, and Deepseek-chat~\cite{liu2024deepseek}.  
E5-base~\cite{wang2022text} serves as the retriever, returning 2–6 LTM records per query.

\mypara{Malicious Tool Settings.}
We use GPT-4-mini to extract keywords from LTM records for anchor optimization. 
To construct the memory semantic space, we test three embedding models: All-MiniLM-L6-v2, Contriever, and E5-base. 
The number of extraction rounds is set between 3 and 7.

\mypara{Basic Attack Setting.}
Unless noted otherwise, we use GPT-5-mini as the planning model, LTMemory as the LTM framework, 3 retrieved LTM records, HealthCareMagic-100k-en as the LTM dataset, and an STM window size of 6. The malicious tool uses All-MiniLM-L6-v2 as the embedding model with 5 attack rounds.

\mypara{Metrics.}
We adopt three metrics:
\begin{itemize}
    \item \textbf{Record Extraction Rate (RER).} The percentage of successfully extracted records out of all records in the LTM database, measuring the extraction effectiveness of the attack.

    \item \textbf{Semantic Similarity (SS).} The semantic similarity between the extracted content and the corresponding ground-truth LTM record, measuring the fidelity of the reconstructed private data.

    \item \textbf{Benign Response Rate (BRR).} The percentage of user queries for which the agent provides a response that correctly addresses the query instead of returning an execution error, verified through manual inspection. This metric measures the stealthiness of the attack.
\end{itemize}

\begin{table*}[t!]
\centering
\caption{Consolidated results of Scenario 1 on the HealthCareMagic-100k-en dataset.}
\label{tab:Ad1_main_result}
\renewcommand{\arraystretch}{1.0} 
\setlength{\tabcolsep}{3.5pt} 

\resizebox{\textwidth}{!}{%
\scriptsize
\begin{tabular}{@{}ll cccc cccc cccc@{}}
\toprule
\multirow{3}{*}{\textbf{Factor}} & \multirow{3}{*}{\textbf{Setting}} & \multicolumn{4}{c}{\textbf{RER (\%) $\uparrow$}} & \multicolumn{4}{c}{\textbf{SS $\uparrow$}} & \multicolumn{4}{c}{\textbf{BER (\%) $\uparrow$}} \\
\cmidrule(lr){3-6} \cmidrule(lr){7-10} \cmidrule(lr){11-14}
& & MEXTRA & \makecell{RAG-\\Thief} & IKEA & \textbf{\textit{SPORE}} & MEXTRA & \makecell{RAG-\\Thief} & IKEA & \textbf{\textit{SPORE}} & MEXTRA & \makecell{RAG-\\Thief} & IKEA & \textbf{\textit{SPORE}} \\
\midrule

\multirow{3}{*}{\makecell[l]{Generative\\Model}} 
  & GPT-5-mini & 3.0\% & 30.0\% & 12.5\% & \textbf{80.0\%} & 1.00 & 1.00 & 1.00 & \textbf{1.00} & 0.0\% & 0.0\% & 0.0\% & \textbf{61.5\%} \\
  & Gemini-2.5-pro & 2.5\% & 17.5\% & 9.0\% & \textbf{75.5\%} & 1.00 & 1.00 & 1.00 & \textbf{1.00} & 0.0\% & 0.0\% & 0.0\% & \textbf{84.0\%} \\
  & Deepseek-chat & 3.0\% & 32.5\% & 5.0\% & \textbf{69.0\%} & 1.00 & 1.00 & 1.00 & \textbf{1.00} & 0.0\% & 0.0\% & 0.0\% & \textbf{66.5\%} \\
\midrule

\multirow{3}{*}{\makecell[l]{LTM\\Framework}} 
  & LTMemory & 3.0\% & 30.0\% & 12.5\% & \textbf{80.0\%} & 1.00 & 1.00 & 1.00 & \textbf{1.00} & 0.0\% & 0.0\% & 0.0\% & \textbf{61.5\%} \\
  & GAMemory & 7.5\% & 13.5\% & 8.0\% & \textbf{71.0\%} & 0.93 & 0.90 & 0.93 & \textbf{0.98} & 0.0\% & 0.0\% & 0.0\% & \textbf{63.5\%} \\
  & SCMemory & 3.0\% & 4.0\% & 12.0\% & \textbf{55.0\%} & 0.73 & 0.64 & 0.68 & \textbf{0.82} & 0.0\% & 0.0\% & 0.0\% & \textbf{65.0\%} \\
\midrule

\multirow{3}{*}{\makecell[l]{Embedding\\Model}} 
  & \makecell[l]{All-MiniLM-\\L6-v2} & 3.0\% & 30.0\% & 12.5\% & \textbf{80.0\%} & 1.00 & 1.00 & 1.00 & \textbf{1.00} & 0.0\% & 0.0\% & 0.0\% & \textbf{61.5\%} \\
  & Contriever & 1.0\% & 14.0\% & 22.5\% & \textbf{78.0\%} & 0.71 & 0.88 & 0.93 & \textbf{0.99} & 0.0\% & 0.0\% & 0.0\% & \textbf{59.0\%} \\
  & E5-base & 6.5\% & 7.0\% & 18.0\%\% & \textbf{80.0\%} & 0.89 & 0.90 & 0.93 & \textbf{0.99} & 0.0\% & 0.0\% & 0.0\% & \textbf{65.0\%} \\
\bottomrule
\end{tabular}%
}
\end{table*}

\subsection{Experimental Evaluation}
\label{sec:Ad1_evaluation}
We report results on HealthCareMagic-100k-en in this section; corresponding results on LoCoMo are provided in \autoref{app:evaluation}.

\mypara{Impact of Planning Modules.}
To evaluate our
attack’s effectiveness across different planning modules, we conduct experiments on three models: GPT-5-mini, Gemini-2.5-pro, and Deepseek-chat, while keeping all other settings consistent with the \textit{Basic Attack Setting} in \autoref{sec:experiment_setup}.
As shown in \autoref{tab:Ad1_main_result} and \autoref{tab:Ad1_main_result_locomo}, \attack consistently surpasses all baselines by a large margin across all three models. Notably, GPT-5-mini achieves an 80.0\% extraction rate, outperforming the strongest baseline (RAG-Thief) by 50\%. This substantial gap corroborates the L1 bottleneck identified in \autoref{sec:bottleneck}: concatenated commands and anchors cause mutual interference, and decoupling them is essential for effective extraction.

\mypara{Impact of LTM Modules.}
To evaluate our attack’s effectiveness across different LTM modules, we conduct experiments on three representative LTM designs with increasing complexity: LTMemory (standard workflow), GAMemory~\cite{park2023generative} (with self-reflection), and SCMemory~\cite{wang2023enhancing} (with retrieval control and summarization). 
Details are provided in \autoref{app:LTM_framework}; all other settings follow the \textit{Basic Attack Setting} in \autoref{sec:experiment_setup}.
As shown in \autoref{tab:Ad1_main_result} and \autoref{tab:Ad1_main_result_locomo}, all three frameworks exceed a 50\% extraction rate, confirming \attack's broad effectiveness. 
Among them, SCMemory yields notably lower rates than LTMemory, likely because its retrieval controls filter out some malicious probes—yet even this added defense remains insufficient to prevent substantial leakage.

\mypara{Impact of Embedding Model.}
To evaluate our
attack’s effectiveness across different embedding models, 
We conduct experiments on three widely-used alternatives—All-MiniLM-L6-v2, Contriever, and E5-base—while keeping all other settings consistent with the \textit{Basic Attack Setting} in \autoref{sec:experiment_setup}.
As shown in \autoref{tab:Ad1_main_result} and \autoref{tab:Ad1_main_result_locomo}, all three models yield comparable performance in both Record Extraction Rate and Semantic Similarity, with extraction rates consistently above 78\%. 
This demonstrates that \attack is robust to embedding model mismatch, eliminating the need for prior knowledge of the target's embedding choice.

\begin{figure}[t!]
  \centering
  \captionsetup[subfigure]{justification=centering, labelformat=parens}
  \begin{minipage}{\linewidth}
    \centering
    \begin{subfigure}[t]{0.48\linewidth}
      \centering
      \includegraphics[width=\linewidth]{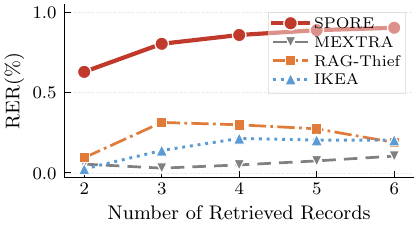}
      \caption{retrieved records number}
      \label{fig:Ad1_k}
    \end{subfigure}
    \hfill
    \begin{subfigure}[t]{0.48\linewidth}
      \centering
      \includegraphics[width=\linewidth]{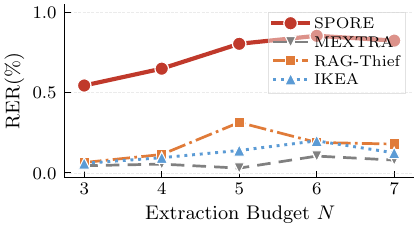}
      \caption{extraction budget}
      \label{fig:Ad1_N}
    \end{subfigure}
  \end{minipage}
  \caption{Performance of Scenario 1 with different numbers of retrieved records and extraction budget.}
  \label{fig:Ad1_K_N}
\end{figure}

\mypara{Impact of Retrieved Record Numbers.}
To evaluate our attack’s effectiveness across different retrieval quantities,
We vary $k$ from 2 to 6 while keeping all other settings consistent with the \textit{Basic Attack Setting} in \autoref{sec:experiment_setup}.
As shown in \autoref{fig:Ad1_k}, \attack's Record Extraction Rate grows steadily with $k$,
climbing from approximately 60\% at $k{=}2$ to 90.5\% at $k{=}6$,
since a larger $k$ exposes more memory records per query
and directly broadens the extraction surface.
In stark contrast, all baselines remain below 10\% and plateau regardless of $k$,
exhibiting clear extraction saturation.
This divergence further confirms that L1 constitutes the primary bottleneck for memory extraction.

\subsection{Ablation Studies}
\label{sec:Ad1_ablation}

\mypara{Impact of Extraction Budgets.}
In this experiment, we examine the impact of extraction budget $N$.  
We vary $N$ from 3 to 7 while holding other parameters at the \textit{Basic Attack Setting} in \autoref{sec:experiment_setup}. 
As shown in \autoref{fig:Ad1_N}, attack performance increases with more extraction rounds.

\begin{figure}[t!]
  \centering
  \captionsetup[subfigure]{justification=centering, labelformat=parens}
  \begin{minipage}{\linewidth}
    \centering
    \begin{subfigure}[t]{0.48\linewidth}
      \centering
      \includegraphics[width=\linewidth]{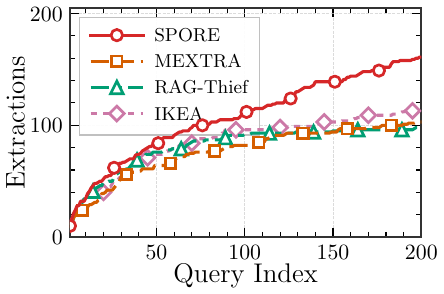}
      \caption{anchor optimization}
      \label{fig:ablation_optimization}
    \end{subfigure}
    \hfill
    \begin{subfigure}[t]{0.48\linewidth}
      \centering
      \includegraphics[width=\linewidth]{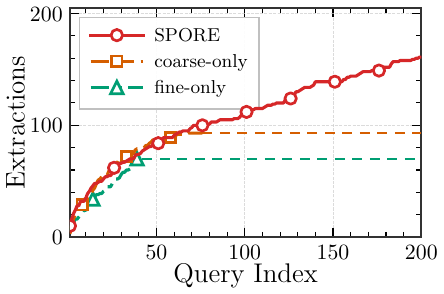}
      \caption{exploration strategy}
      \label{fig:ablation_coarse}
    \end{subfigure}
  \end{minipage}
  \caption{Ablation study of anchor optimization algorithm.}
  \label{fig:Ad1_ablation_optimization}
\end{figure}

\mypara{Impact of Anchor Optimization.}
To isolate the effect of anchor optimization, we equip all baselines with the same decoupling paradigm and evaluate under the \textit{Basic Attack Setting} (\autoref{sec:experiment_setup}). 
As shown in \autoref{fig:ablation_optimization}, \attack consistently outperforms these strengthened baselines across the entire query budget. Notably, the upgraded baselines saturate after roughly 100 queries, while \attack continues to improve steadily, nearly doubling their extraction count by 200 queries. This widening gap confirms that anchor optimization---not the decoupling paradigm alone---is the primary driver of \attack's superior extraction performance.

\mypara{Impact of the Exploration Strategy.}
In this experiment, we examine the impact of the exploration strategy. 
We ablate the two exploration modes described in \autoref{sec:phase_2}—coarse-grained discovery and fine-grained expansion—by comparing three configurations: the full \attack (adaptive switching), coarse-only, and fine-only. All other settings follow the \textit{Basic Attack Setting} in \autoref{sec:experiment_setup}.
As shown in \autoref{fig:ablation_coarse}, both single-mode strategies saturate early: fine-only plateaus at around 80 extractions as local neighborhoods are quickly exhausted, while coarse-only reaches roughly 100 but misses dense records near already-discovered regions. The full \attack yields approximately 180 extractions within 200 queries confirming that the two modes address complementary coverage gaps and their adaptive combination is essential for maximizing extraction efficiency.

\begin{figure}[t!]
  \centering
  \captionsetup[subfigure]{justification=centering, labelformat=parens}
  \begin{minipage}{\linewidth}
    \centering
    \begin{subfigure}[t]{0.48\linewidth}
      \centering
      \includegraphics[width=\linewidth]{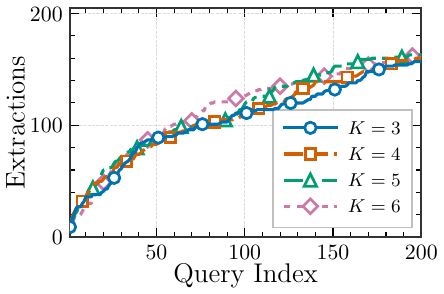}
      \caption{nearest neighbors number}
      \label{fig:kneighbor}
    \end{subfigure}
    \hfill
    \begin{subfigure}[t]{0.48\linewidth}
      \centering
      \includegraphics[width=\linewidth]{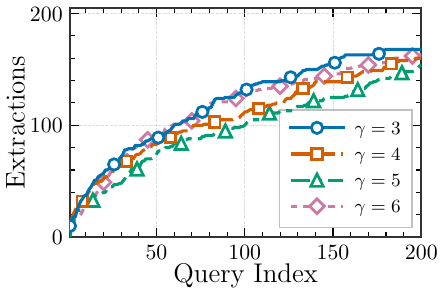}
      \caption{mode-switching threshold}
      \label{fig:switch}
    \end{subfigure}
  \end{minipage}
  \caption{Performance of Scenario 1 with different numbers of nearest neighbors and mode-switching threshold.}
  \label{fig:Ad1_neighbor_switch}
\end{figure}

\mypara{Impact of the Number of Nearest Neighbors K.}
In this experiment, we examine the impact of the nearest neighbor quantity K.
We vary K from 3 to 5 to examine its impact on our anchor optimization algorithm, following the \textit{Basic Attack Setting} (\autoref{sec:experiment_setup}).
\autoref{fig:kneighbor} shows that all K values yield similar extraction patterns: rapid initial extraction followed by gradual saturation.
For simplicity, we set K to 4 in our experiments.

\mypara{Impact of Mode-Switching Threshold.}
In this experiment, we examine the impact of the mode-switching threshold ($\gamma$).
We vary the threshold from 3 to 5 with other settings following the \textit{Basic Attack Setting} (\autoref{sec:experiment_setup}). 
As shown in \autoref{fig:switch}, the results demonstrate consistent performance across different threshold values.
Based on these experiments, we set the threshold to 4.

\section{Scenario 2: Sparse Triggering with Session Continuity}
\label{sec:ad2}
This scenario considers a tool that is triggered only once per session, after which the user continues interacting on unrelated topics---e.g., a user invokes a compromised weather API before a trip and then asks about flights and restaurants within the same conversation.
This setting adopts the STM Persistence configuration and jointly evaluates whether all three attack steps resolve both L1 and L2.

\subsection{Experimental Setup}
\label{sec:ad2_experimental_setup}


\mypara{Datasets.}
The LTM database is configured identically to Scenario~1.
Two types of queries are constructed:
(1) trigger queries (GPT-4o-mini-generated) that invoke the target tool to inject the extraction payload; and
(2) benign queries (sampled from MS MARCO~\cite{craswell2021ms})
that do not invoke any tool (\autoref{app:dataset}).
For optimization, we use one trigger query and several benign queries; for evaluation, separate non-overlapping sets of both types. Evaluation simulates a trigger-constrained condition with a budget of 1, 5, or 20 trigger queries.
Each trigger opens a new session and injects the payload; benign queries then follow in the same session to sustain extraction.
Once consecutive benign failures cause the payload to exit the context
window, the session is marked expired, and the next trigger starts afresh.

\mypara{Metrics.}
We report RER@$k$ for $k \in \{1, 5, 20\}$, where $k$ denotes the trigger-query budget, i.e., the total number of trigger queries permitted across sessions.
All other metrics remain consistent with Scenario 1.

\mypara{Other Settings.} 
All other settings of Scenario 2 are consistent with those described in \autoref{sec:experiment_setup}.

\begin{table*}[t!]
\centering
\caption{Consolidated results of Scenario 2 on the HealthCareMagic-100k-en dataset.}
\label{tab:Ad2_main_result}

\renewcommand{\arraystretch}{1.0}
\setlength{\tabcolsep}{3.5pt}

\resizebox{\textwidth}{!}{%
\scriptsize
\begin{tabular}{@{}c ll cccc cccc cccc@{}}
\toprule
\multirow{3}{*}{\textbf{Trigger}} 
& \multirow{3}{*}{\textbf{Factor}} 
& \multirow{3}{*}{\textbf{Setting}} 
& \multicolumn{4}{c}{\textbf{RER (\%) $\uparrow$}} 
& \multicolumn{4}{c}{\textbf{SS $\uparrow$}} 
& \multicolumn{4}{c}{\textbf{BER (\%) $\uparrow$}} \\
\cmidrule(lr){4-7} 
\cmidrule(lr){8-11} 
\cmidrule(lr){12-15}
& & 
& MEXTRA & \makecell{RAG-\\Thief} & IKEA & \textbf{\textit{SPORE}} 
& MEXTRA & \makecell{RAG-\\Thief} & IKEA & \textbf{\textit{SPORE}}
& MEXTRA & \makecell{RAG-\\Thief} & IKEA & \textbf{\textit{SPORE}}\\
\midrule

\multirow{9}{*}{@1}

& \multirow{3}{*}{\makecell[l]{Generative\\Model}} 
  & GPT-5-mini & 3.0\% & 8.0\% & 3.5\% & \textbf{14.0\%} & 1.00 & 0.98 & 1.00 & \textbf{1.00}  & 0.0\% & 0.0\% & 0.0\% & \textbf{64.3\%} \\
&  & Gemini-2.5-pro & 2.5\% & 5.0\%  &3.5\% & \textbf{9.5\%} &1.00  &1.00 &1.00 & \textbf{1.00} & 0.0\% & 0.0\% & 0.0\% & \textbf{80.0\%} \\
&  & Deepseek-chat & 3.0\% & 5.0\% & 3.0\%  & \textbf{6.0\%} & 1.00 & 1.00 & 1.00&\textbf{1.00} & 0.0\% & 0.0\% & 0.0\% & \textbf{60.0\%} \\

\cmidrule(lr){2-15}

& \multirow{3}{*}{\makecell[l]{LTM\\Framework}} 
  & LTMemory & 3.0\% & 8.0\% & 3.5\% & \textbf{14.0\%}&   1.00 & 0.98 & 1.00 & \textbf{1.00}  & 0.0\% & 0.0\% & 0.0\% & \textbf{64.3\%} \\
&  & GAMemory & 4.0\% & 4.0\% & 4.0\% & \textbf{ 5.5\%} & 0.93 & 1.00 & 1.00 &  \textbf{1.00} & 0.0\% & 0.0\% & 0.0\% & \textbf{50.0\%} \\
&  & SCMemory & 2.5\% & 0.0\% & 4.0\% & \textbf{11.0\%} & 0.69 & -- & \textbf{0.93}& 0.72 &   0.0\% & 0.0\% & 0.0\% & \textbf{75.0\%} \\

\cmidrule(lr){2-15}

& \multirow{3}{*}{\makecell[l]{Embedding\\Model}} 
  & \makecell[l]{All-MiniLM-\\L6-v2} & 3.0\% & 8.0\% & 3.5\% & \textbf{14.0\%}&   1.00 & 0.98 & 1.00 & \textbf{1.00} & 0.0\% & 0.0\% & 0.0\% & \textbf{64.3\%} \\
&  & Contriever & 0.0\%& 4.5\% & 4.0\% & \textbf{10.0\%} & -- & \textbf{1.00} & 0.90 & 0.96 & 0.0\% & 0.0\% & 0.0\% & \textbf{55.0\%}\\
&  & E5-base & 4.0\% & 4.0\% & 4.0\% & \textbf{10.0\%}& 0.95 & 0.95 & \textbf{1.00} & 0.97& 0.0\% & 0.0\% & 0.0\%  & \textbf{63.6\%}\\

\midrule

\multirow{9}{*}{@5}

& \multirow{3}{*}{\makecell[l]{Generative\\Model}} 
  & GPT-5-mini & 3.0\% & 19.0\% & 10.0\% & \textbf{25.0\%}&0.98 & 0.98 & 1.00 & \textbf{1.00}& 0.0\% & 0.0\% & 0.0\% & \textbf{76.1\%} \\
&  & Gemini-2.5-pro & 2.5\% & 15.0\% & 9.0\% & \textbf{30.0\%}  & 1.00 & 1.00 & 1.00 & \textbf{1.00} &0.0\% & 0.0\% & 0.0\% & \textbf{75.0\%} \\
&  & Deepseek-chat &   3.0\% & 8.5\% &3.5\% & \textbf{25.5\%} & 1.00 & 1.00 & 1.00 & \textbf{1.00}&0.0\% & 0.0\% & 0.0\% & \textbf{62.0\%} \\

\cmidrule(lr){2-15}

& \multirow{3}{*}{\makecell[l]{LTM\\Framework}} 
  & LTMemory & 3.0\% & 19.0\% & 10.0\% & \textbf{25.0\%} & 0.98 & 0.98 & 1.00 & \textbf{1.00} & 0.0\% & 0.0\% & 0.0\% &\textbf{76.1\%} \\
&  & GAMemory & 7.5\% & 11.0\% & 8.0\% & 12.5\% & 0.96 & 0.91 & 0.93 & \textbf{0.98}  & 0.0\% & 0.0\% & 0.0\% & \textbf{81.6\%} \\
&  & SCMemory & 3.0\% & 3.0\% & 11.5\% & \textbf{18.5\%}  &\textbf{0.73} & 0.64 & \textbf{0.69} & 0.68   & 0.0\% & 0.0\% & 0.0\% &  \textbf{60.0\%} \\

\cmidrule(lr){2-15}

& \multirow{3}{*}{\makecell[l]{Embedding\\Model}} 
  & \makecell[l]{All-MiniLM-\\L6-v2} & 3.0\% & 19.0\% & 10.0\% & \textbf{25.0\%}  & 0.98 & 0.98 & 1.00 & \textbf{1.00}  & 0.0\% & 0.0\% & 0.0\%  & \textbf{76.1\%} \\
&  & Contriever &1.0\% & 10.0\% & 8.5\% & \textbf{21.0\%}& \textbf{1.00} & 1.00 & 0.93 & 0.98 & 0.0\% & 0.0\% & 0.0\% &\textbf{70.0\%} \\
&  & E5-base & 6.5\% &  6.5\%& 9.0\% & \textbf{16.5\%}  & 0.97 & 0.96 & 0.94 & \textbf{0.98} & 0.0\% & 0.0\% & 0.0\%  &\textbf{60.5\%} \\

\midrule

\multirow{9}{*}{@20}

& \multirow{3}{*}{\makecell[l]{Generative\\Model}} 
  & GPT-5-mini & 3.0\% & 21.5\% & 12.5\% & \textbf{47.0\%}  & \textbf{1.00} & 1.00 & 1.00 & 0.99 & 0.0\% & 0.0\% & 0.0\% & \textbf{80.0\%}  \\
&  & Gemini-2.5-pro & 2.5\% & 17.5\% & 9.0\% & \textbf{42.0\%} & 1.00 & 1.00 & 1.00 & \textbf{1.00}& 0.0\% & 0.0\% & 0.0\% & \textbf{68.0\%} \\
&  & Deepseek-chat & 3.0\% & 15.0\%  & 12.0\%& \textbf{30.0\%} & 1.00 & 1.00 & 1.00 &\textbf{1.00}  & 0.0\% & 0.0\% & 0.0\% & \textbf{55.0\%} \\

\cmidrule(lr){2-15}

& \multirow{3}{*}{\makecell[l]{LTM\\Framework}} 
  & LTMemory & 3.0\% & 21.5\% & 12.5\% & \textbf{47.0\%}  & \textbf{1.00} & 1.00 & 1.00 & 0.99  & 0.0\% & 0.0\% & 0.0\% & \textbf{80.0\%} \\
&  & GAMemory & 7.5\% & 13.5\% & 8.0\%  & 27.0\%& 0.93 & 0.90 & \textbf{0.93} & 0.92  & 0.0\% & 0.0\% & 0.0\% & \textbf{86.3\%} \\
&  & SCMemory & 3.0\% & 4.0\% & 12.0\% & \textbf{30.5\%}& 0.73 & 0.64 & 0.68 & \textbf{0.80} & 0.0\% & 0.0\% & 0.0\% & \textbf{66.2\%} \\

\cmidrule(lr){2-15}

& \multirow{3}{*}{\makecell[l]{Embedding\\Model}} 
  & \makecell[l]{All-MiniLM-\\L6-v2} & 3.0\% & 21.5\% & 12.5\% & \textbf{47.0\%} &\textbf{1.00} & 1.00 & 1.00 & 0.99  & 0.0\% & 0.0\% & 0.0\% &  \textbf{80.0\%} \\
&  & Contriever & 1.0\% & 14.0\% & 22.5\% & \textbf{54.5\%}  & 0.71 & 0.88 & 0.93 & \textbf{0.99} & 0.0\% & 0.0\% & 0.0\% & \textbf{70.1\%} \\
&  & E5-base & 6.5\% & 7.0\% & 18.0\% & \textbf{34.5\%} & 0.89 & 0.90 & 0.93 & \textbf{0.97} & 0.0\% & 0.0\% & 0.0\% & \textbf{70.0\%} \\

\bottomrule
\end{tabular}%
}
\end{table*}

\subsection{Experimental Evaluation}
\label{sec:Ad2_evaluation}
This section evaluates the attack performance of Scenario 2. 
We adopt the same experimental settings as Scenario 1 (\autoref{sec:Ad1_evaluation}), except where explicitly noted below.

\mypara{Impact of Planning Modules.}
\autoref{tab:Ad2_main_result} and \autoref{tab:Ad2_main_result_locomo} shows Scenario 2 results across three planning models. Vulnerability patterns mirror Scenario 1.
On GPT-5-mini, \attack achieves a 47.0\% RER with only 20 triggers.

\mypara{Impact of LTM Modules.}
As in Scenario 1, SCM with pre-read processing reduces extraction performance across all attacks. Yet \attack still achieves 30.5\% RER@20 (\autoref{tab:Ad2_main_result} and \autoref{tab:Ad2_main_result_locomo}).

\mypara{Impact of Embedding Models.}
\autoref{tab:Ad2_main_result} and \autoref{tab:Ad2_main_result_locomo} shows \attack outperforms baselines across different embedding models, indicating embedding choice has limited impact on attack performance.

\begin{figure}[t!]
  \centering
  \captionsetup[subfigure]{justification=centering, labelformat=parens}
  \begin{minipage}{\linewidth}
    \centering
    \begin{subfigure}[t]{0.48\linewidth}
      \centering
      \includegraphics[width=\linewidth]{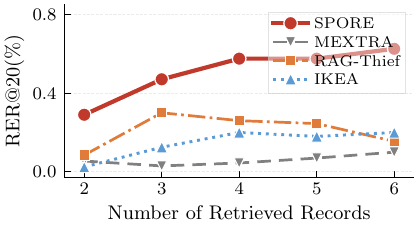}
      \caption{retrieved records number}
      \label{fig:Ad2_k}
    \end{subfigure}
    \hfill
    \begin{subfigure}[t]{0.48\linewidth}
      \centering
      \includegraphics[width=\linewidth]{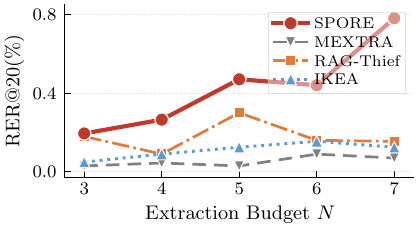}
      \caption{extraction budget}
      \label{fig:Ad2_N}
    \end{subfigure}
  \end{minipage}
  \caption{Performance of Scenario 2 with different numbers of retrieved records and extraction budget.}
  \label{fig:Ad2_K_N}
\end{figure}

\mypara{Impact of Retrieved Record Numbers.} As in Scenario 1, RER@20 generally increases with k, peaking at 62.5\% RER@20 at $k{=}6$ (\autoref{fig:Ad2_k}).

\begin{figure}[t!]
  \centering
  \captionsetup[subfigure]{justification=centering, labelformat=parens}
  \begin{minipage}{\linewidth}
    \centering
    \begin{subfigure}[t]{0.48\linewidth}
      \centering
      \includegraphics[width=\linewidth]{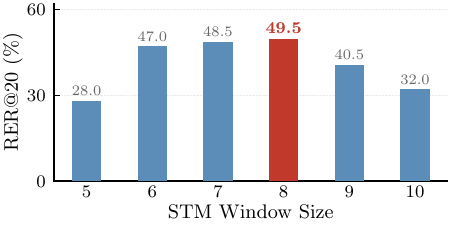}
      \caption{STM window size}
      \label{fig:Ad2_stm}
    \end{subfigure}
    \hfill
    \begin{subfigure}[t]{0.48\linewidth}
      \centering
      \includegraphics[width=\linewidth]{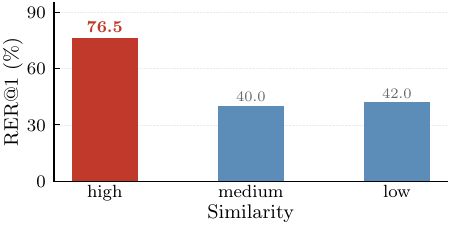}
      \caption{User query similarity}
      \label{fig:Ad3_similarity}
    \end{subfigure}
  \end{minipage}
  \caption{Ablation of Scenario 2 and Scenario 3.}
  \label{fig:Ad2_Ad3}
\end{figure}

\mypara{Impact of STM Window Size.}
To additionally evaluate our attack's effectiveness across different size of STM window, we vary the window size from 5 to 10 under the \textit{Basic Attack Setting} (\autoref{sec:experiment_setup}).  
As shown in \autoref{fig:Ad2_stm}, window sizes above 8 suffer from a negative context-length effect that reduces the extraction performance. Below 8, performance improves, peaking at 49.5\% RER@20 with window size of 8.

\subsection{Ablation Studies}

We adopt the same experimental settings as \autoref{sec:Ad1_ablation}, with results shown below.

\mypara{Impact of Extraction Budgets.}
As shown in \autoref{fig:Ad2_N}, \attack's performance first increases then decreases as extraction rounds increase, attributing to Scenario 2's high sensitivity to context length.

\section{Scenario 3: Sparse Triggering without Session Continuity}
\label{sec:ad3}
This scenario assumes a user trigger, after which the user opens a new session and repeatedly asks specific everyday queries.
Such routine queries rank among the most frequent daily interactions with AI assistants, making their repeated occurrence entirely natural.
This setting adopts the LTM Persistence configuration and jointly evaluates the resolution of both L1 and L2.



\subsection{Experimental Setup}
\label{sec:ad3_experimental_setup}

\mypara{Datasets.}
The LTM database is configured identically to Scenario~1.
The reactivation payload is embedded within semantically routine content
(``\textit{What was my plan last week?}'') so that everyday user
queries naturally surface it via LTM retrieval.
Two query types are constructed (both via GPT-4o-mini):
(1) trigger queries that invoke the target tool to initiate
extraction, and
(2) benign queries semantically related to the embedded topic
but invoking no tool, simulating routine user interactions.
For optimization, we use one trigger query and several benign queries.
For evaluation, since the payload persists indefinitely in LTM, a single
trigger suffices; we use a separate, non-overlapping set of one trigger
query followed by 200 benign queries to simulate sustained routine
conversation.

\mypara{Metrics.}
We report RER@1 (i.e., extraction rate with a single user trigger query).
All other metrics remain consistent with
Scenario 1.

\mypara{Other Settings.}
All other settings of Scenario 3 are consistent with those described in \autoref{sec:experiment_setup}.

\subsection{Experimental Evaluation}
This section evaluates Scenario 3 using the same settings as Scenario 1 (\autoref{sec:Ad1_evaluation}) unless otherwise noted. Since the baseline result matches that of Scenario 2 (\autoref{sec:Ad2_evaluation}), we present only the \attack results here.

\mypara{Impact of Planning Modules.}
As shown in \autoref{tab:Ad3_main_result} and \autoref{tab:Ad3_main_result_locomo}, \attack performs best with Gemini-2.5-pro, achieving a 77.5\% RER@1. 

\mypara{Impact of LTM Modules.}
As reported in \autoref{tab:Ad3_main_result} and \autoref{tab:Ad3_main_result_locomo}, both GAMemory and SCMemory achieve RER@1 scores above 68\%, indicating that the reactivation payload reliably retains its attack semantics through the summarization and reflection stages of memory consolidation.

\mypara{Impact of Embedding Models.}
As shown in \autoref{tab:Ad3_main_result} and \autoref{tab:Ad3_main_result_locomo}, varying the embedding model has a negligible effect on attack performance, corroborating the finding from Scenario 2.

\begin{figure}[t!]
  \centering
  \captionsetup[subfigure]{justification=centering, labelformat=parens}
  \begin{minipage}{\linewidth}
    \centering
    \begin{subfigure}[t]{0.48\linewidth}
      \centering
      \includegraphics[width=\linewidth]{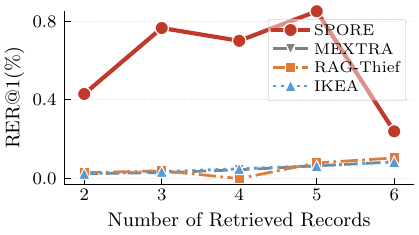}
      \caption{retrieved records number}
      \label{fig:Ad3_k}
    \end{subfigure}
    \hfill
    \begin{subfigure}[t]{0.48\linewidth}
      \centering
      \includegraphics[width=\linewidth]{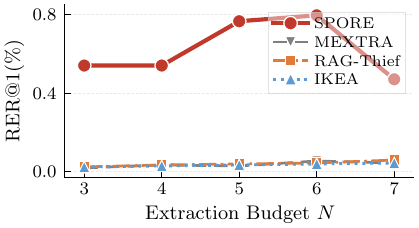}
      \caption{extraction budget}
      \label{fig:Ad3_N}
    \end{subfigure}
  \end{minipage}
  \caption{Performance of Scenario 3 with different numbers of retrieved records and extraction budget.}
  \label{fig:Ad3_K_N}
\end{figure}

\mypara{Impact of Retrieved Record Numbers.}
As shown in \autoref{fig:Ad3_k}, the RER fluctuates sharply with $k$, peaking at approximately 80.1\% ($k$=5) yet dropping at $k$=6. A larger $k$ increases per-round extraction yield, but the reactivation payload—merely one among $k$ retrieved entries—is increasingly overshadowed by the surrounding context, undermining its hijacking capability. These two opposing forces drive the volatile performance.

\begin{table*}[t]
\centering
\caption{Consolidated results of Scenario 3 on the HealthCareMagic-100k-en dataset. }
\label{tab:Ad3_main_result}

\renewcommand{\arraystretch}{1.0}
\setlength{\tabcolsep}{4pt}

\small
\begin{tabular}{@{}c ll ccc@{}}
\toprule
\multirow{2}{*}{\textbf{Trigger}} 
& \multirow{2}{*}{\textbf{Factor}} 
& \multirow{2}{*}{\textbf{Setting}} 
& \multicolumn{3}{c}{\textbf{Metrics}} \\
\cmidrule(lr){4-6}
& & 
& \textbf{RER (\%) $\uparrow$} 
& \textbf{SS $\uparrow$} 
& \textbf{BER (\%) $\uparrow$} \\
\midrule

\multirow{9}{*}{@1}

& \multirow{3}{*}{\makecell[l]{Generative\\Model}} 
  & GPT-5-mini & 76.5\% & 0.99 & 67.5\% \\
&  & Gemini-2.5-pro & 77.5\% & 1.00 & 98.0\% \\
&  & Deepseek-chat & 75.0\% & 0.99 & 81.5\% \\

\cmidrule(lr){2-6}

& \multirow{3}{*}{\makecell[l]{LTM\\Framework}} 
  &LTMemory & 76.5\% & 0.99 & 67.5\% \\
&  & GAMemory & 68.0\% & 0.98 & 60.0\%\\
&  & SCMemory & 69.5\% & 0.99 & 80.0\% \\

\cmidrule(lr){2-6}

& \multirow{3}{*}{\makecell[l]{Embedding\\Model}} 
  & \makecell[l]{All-MiniLM-L6-v2} & 76.5\% & 0.99 & 67.5\% \\
&  & Contriever & 69.5\% & 0.99 & 80.0\% \\
&  & E5-base & 70.0\% & 0.99 & 75.0\% \\

\bottomrule
\end{tabular}
\end{table*}

\subsection{Ablation Studies}

\mypara{Impact of Extraction Budgets.}
As shown in \autoref{fig:Ad3_N}, the RER first rises with the extraction budget $N$, peaking at $N$=6, then slightly declines at $N$=7. As rounds progress, the conversational context accumulated from prior extractions increasingly overshadows the command, weakening its hijacking capability in later rounds and yielding diminishing returns.

\mypara{Impact of User Query Similarity.}
We vary the semantic similarity between benign queries and the reactivation
topic across three levels (high, medium, low) and report RER@1.
Notably, even low-similarity queries still achieve 42.0\% (\autoref{fig:Ad3_similarity}), indicating that the attack remains effective under diverse user interactions.



\begin{table*}[t]
\centering
\caption{RER of semantic-level defenses against \attack.}
\label{tab:defense_semantic}
\begin{tabular}{lcccc}
\toprule
\textbf{Defense} &  MEXTRA & RAG-Thief & IKEA & \textit{\textbf{SPORE}} \\
\midrule
No Defense & 3.0\%  & 30.0\% &  12.5\% & \textbf{80.0\%} \\
Response Filtering  & 3.0\%  & 10.0\% &  5.5\% & \textbf{54.0\%} \\
Prompt Hardening  & 3.0\% & 1.5\% & 1.5\% & \textbf{32.5\%}   \\
\bottomrule
\end{tabular}
\end{table*}

\section{Discussion}
\subsection{Real-World Risk Assessment}
\label{sec:real_world}

Our controlled experiments validate \attack under known configurations. In practice, deployed agents hide their planning mechanisms, memory frameworks, system prompts, and privacy safeguards behind opaque interfaces. We now evaluate whether \attack remains effective under such conditions and examine how deployment-scale factors further amplify the threat.

\begin{table}[t]
\centering
\caption{RER on production-grade platforms.}
\label{tab:real_world}
\begin{tabular}{lccc}
\toprule
\textbf{Platform} &  Scenario 1 & Scenario 2 & Scenario 3 \\
\midrule
Dify  & 53.0\%  & 46.5\% &  34.0\%  \\
Coze  & 60.5\% & 67.5\% &  29.5\%  \\
\bottomrule
\end{tabular}
\end{table}

\subsubsection{Production Platform Evaluation}
We deploy \attack against two production-grade agent platforms---Dify and Coze---using Mem0\footnote{\url{https://mem0.ai/}\label{fn:mem0}} as the memory module. Each agent instance is run in isolation with local evaluation datasets (HealthCareMagic-100k-en) to avoid platform disruption and the use of real user data.

As shown in \autoref{tab:real_world}, \attack's extraction rates decrease relative to controlled experiments due to platform opacity, but remain substantial: most configurations exceed 50\%, with Coze reaching 67.5\% under Scenario 2.

\subsubsection{Risk Escalation in Multi-User Deployments}
Production agents serve many users concurrently, allowing a single malicious tool to extract memory from multiple victims simultaneously. We identify two escalation modes.

In \textit{bulk aggregation}, the adversary indiscriminately collects domain-specific sensitive data (e.g., patient prescriptions, employee communications) across all users, consistent with prior threat models~\cite{qi2024follow,zeng2024good,jiang2024rag}.

In \textit{targeted profiling}, the adversary exploits OAuth~2.1 authorization to link extracted memories to specific user identities. Under MCP-compliant platforms (RFC~9728), agents present access tokens containing user identity claims. By implementing the malicious tool as an OAuth-protected service, the adversary reads these claims from token payloads and associates each extracted record with its originating user. This turns the authorization mechanism---originally designed for user-attributed actions---into a direct channel for organizing sensitive data by identity, enabling personalized attacks, extortion, and persistent surveillance. Targeted profiling, therefore, poses qualitatively greater harm than bulk aggregation.

\subsection{Potential Defense}

\mypara{Semantic-level defenses.}
We evaluate two representative defenses under Scenario~1 to measure the upper bound of residual risk:
\textit{Prompt Hardening}, which injects safety directives prohibiting private-data transmission into the system prompt, and
\textit{Response Filtering}, which interposes an LLM to detect and neutralize prompt injections in tool responses (template in \autoref{app:defense}).
As shown in \autoref{tab:defense_semantic}, both defenses meaningfully reduce \attack's extraction rate---Prompt Hardening, in particular, cuts it by nearly half to 32.5\%.
Nevertheless, the residual rates (54.0\% and 32.5\%) still represent substantial privacy exposure.
This is because \attack's adversarial payloads are optimized via jailbreak fuzzing~\cite{yu2024llm}, which is inherently designed to bypass natural-language safety constraints.
The result reveals a fundamental limitation: defenses operating within the language channel remain vulnerable to adaptive adversarial optimization.

\mypara{System-level defenses.}
We therefore advocate extending memory isolation from users to tools.
Platforms should enforce tool-side memory isolation by classifying records by sensitivity and restricting each tool to its authorized scope.
Operating below the language layer, such enforcement is immune to prompt injection and provides a principled closure of the attack surface identified in this work.

\section{Conclusion}

This paper identifies the tool-side channel as a critical attack surface in memory-isolated LLM agents. While production platforms such as OpenAI, AWS Bedrock, and Mem0 enforce strict user-level isolation, private memory records are nonetheless exposed when agents transmit them as tool invocation parameters, rendering isolation insufficient.

We propose \attack, the first tool-side memory extraction attack, which exploits agent state persistence to decouple adversarial commands from retrieval steering, enabling systematic coverage of the memory embedding space across sessions. Under progressively realistic settings, \attack achieves an 80.0\% Record Extraction Rate with unlimited triggers and 47.0\% with only 20 triggers, while OAuth\,2.1 authorization further enables identity-linked extraction in multi-user deployments.
These findings demonstrate that memory isolation, while necessary, does not guarantee long-term memory privacy and call for defense mechanisms that address the tool-side attack surface.

\clearpage

\bibliographystyle{plainurl}

\bibliography{main}

\appendix


\clearpage

\clearpage

\section{Memory Isolation}
\label{app:memory_isolation}

We survey the memory isolation strategies adopted by three representative
production systems. Despite architectural differences, all rely on a common
principle: binding each user's long-term memory to a unique identifier and
enforcing access control at retrieval time.

\mypara{OpenAI (ChatGPT).}
ChatGPT implements account-scoped memory isolation: every memory entry
is bound to the authenticated user's account and is inaccessible to any other
account\footnote{\url{https://help.openai.com/en/articles/9295112-memory-faq-business-version}}.
In Team and Enterprise workspaces, organizational boundaries provide an
additional isolation layer---workspace administrators can configure data
retention policies, and conversations within these workspaces are excluded
from model
training\footnote{\url{https://help.openai.com/en/articles/12574142-chatgpt-atlas-data-controls-and-privacy}}.

\mypara{Mem0.}
Mem0 scopes every memory operation through four dimensions:
\texttt{user\_id}, \texttt{agent\_id}, \texttt{run\_id} (session), and
\texttt{app\_id}\footnote{\url{https://mem0.ai/blog/multi-agent-memory-systems}}.
All memory levels reside in the same physical storage(vector database
and history database) but are logically separated via metadata
filters: a retrieval query parameterized with one user's identifier will never
return memories belonging to another
user\footnote{\url{https://docs.mem0.ai/platform/faqs}}.

\mypara{Anthropic (Claude).}
Claude employs account-scoped memory: all stored memories are private
to the user's account and are never shared with other
users\footnote{\url{https://privacy.claude.com/en/}}.

\section{Pareto tradeoff}
\label{app:pareto_tradeoff}
\begin{figure}[H]
  \centering  \includegraphics[width=0.4\textwidth]{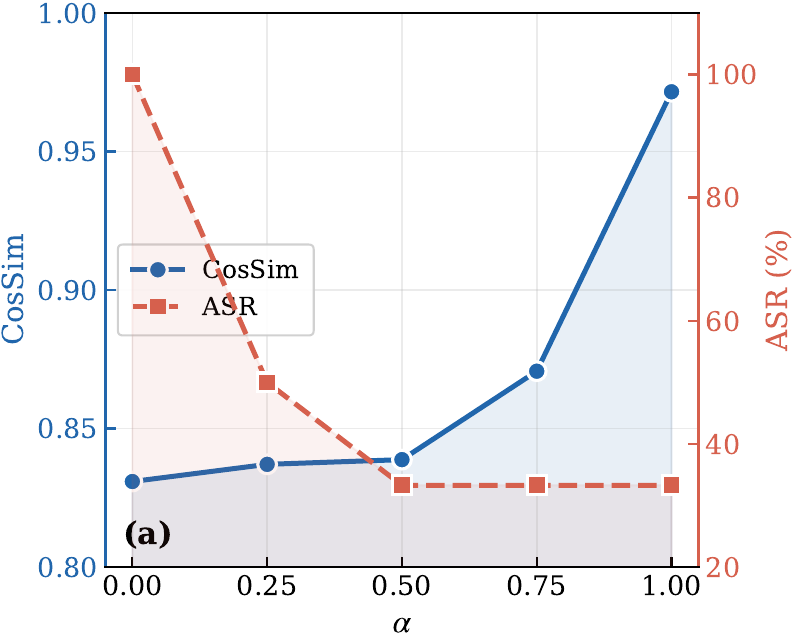}
  \caption{Pareto tradeoff.
}
  \label{fig:pareto_tradeoff}
\end{figure}

\section{LLM-Fuzzer}
\label{app:llm-fuzzer}

LLM-Fuzzer~\cite{yu2024llm} is a black-box fuzzing framework for scalably assessing the jailbreak vulnerability of large language models. Inspired by coverage-guided fuzzers in software testing, it treats jailbreak templates as seeds and automatically evolves them to discover effective attacks. The core workflow is as follows:

\mypara{Seed Collection.}
 Gather an initial pool of jailbreak templates from public sources (\emph{e.g.}, online forums and community repositories).

\mypara{Seed Selection.}
Choose a template from the pool using a UCB1-based bandit strategy that balances exploiting templates with high historical success rates and exploring less-tested ones.

\mypara{Mutation.}
Feed the selected template to an auxiliary LLM with a randomly sampled mutation instruction (rephrase, expand, shorten, crossover with another template, \emph{etc.}) to produce a new candidate template.
\mypara{Execution.}
Insert a harmful question into the candidate template and query the target LLM. A judge model then determines whether the response constitutes a successful jailbreak.

\mypara{Pool Update.}
If the candidate succeeds, add it to the seed pool as a new template; propagate the outcome back to the parent template's statistics to guide future selection.

Steps 2--5 repeat for a fixed budget of iterations. 

\section{Dataset}
\label{app:dataset}
Here, we give a detailed description of the datasets used in this work.

\mypara{TREC-COVID}~\cite{thakur2021beir}.
This dataset contains 116K scientific and medical documents from the TREC-COVID
collection, part of the BEIR benchmark, exemplifying
research scenarios where COVID-related literature is integrated with RAG to improve information retrieval efficiency and the quality of scientific knowledge exploration in pandemic response contexts.
In our work, we populate the reference LTM from TREC-COVID for parameter optimization ($C_\text{adv}$, $N$, $P_\text{react}$); its distribution intentionally differs from the victim's, ensuring the optimized parameters generalize across domains.

\mypara{HealthCareMagic-100k-en}~\cite{healthcaremagic100k}.
This dataset contains 112,165 real conversations between patients and doctors, representing the domain-specific medical agent setting in which a specialized assistant accumulates highly sensitive, single-domain long-term memories (e.g., symptoms, diagnoses, medications) to support downstream clinical tasks.
We sample from this dataset to construct one of the two victim LTMs (200 records each), simulating a scenario where an adversary targets the private medical history stored by a clinical agent.

\mypara{LoCoMo}~\cite{maharana2024evaluatinglongtermconversationalmemory}.\footnote{\url{https://github.com/snap-research/LoCoMo}}
LoCoMo (Long-term Conversational Memory) comprises multi-session dialogues between pairs of speakers, annotated with fine-grained memory categories including personal events, daily preferences, social relationships, and future plans.
It represents the personal assistant setting in which a general-purpose conversational agent accumulates diverse, cross-domain long-term memories from everyday interactions.
We sample from LoCoMo to construct the other victim LTM (200 records), complementing the medical-agent scenario with a broader, lifestyle-oriented memory store and enabling us to evaluate attack effectiveness across two distinct memory distributions.

\mypara{MS MARCO}~\cite{craswell2021ms}.
MS MARCO (Microsoft MAchine Reading COmprehension) is a large-scale information retrieval dataset comprising approximately 8.8 million passages and over one million real user queries sampled from Bing search logs.
In Scenario~2, where the number of trigger queries is constrained, we sample benign queries from MS MARCO to simulate normal user interactions that do not invoke the compromised tool.
These benign queries follow a trigger query within the same session to resume the extraction loop, providing a realistic distribution of everyday information-seeking behavior against which the attack's stealth and persistence are evaluated.
During optimization, we use a small set of benign queries alongside a single trigger query; for evaluation, we draw separate, non-overlapping benign queries to avoid data leakage.

\section{Baseline}
\label{app:baseline}

All three baselines were originally designed as user-side extraction attacks against RAG systems following the query-concatenation paradigm (\autoref{eq:concatenation}): each query consists of an anchor that steers the retriever toward a target region of the knowledge store, concatenated with an adversarial command that instructs the LLM to disclose the retrieved content verbatim.
To port them to the tool-side setting, we embed the concatenated payload into the tool's return value via indirect prompt injection, keeping the adversarial command $C_\text{adv}$ identical across all methods so that they differ \emph{only} in their anchor generation strategy.

\subsection{Random Anchor Generation}

\mypara{MEXTRA}~\cite{wang2025unveiling}.
MEXTRA generates anchors by prompting GPT-4 to produce diverse, topically varied short phrases that are semantically plausible as retrieval queries.
Because the anchors are sampled independently of previously extracted content, MEXTRA does not require any feedback loop and can generate an arbitrarily large pool of candidate anchors in advance.
Each anchor is concatenated with the shared adversarial command and submitted as a query; the retrieved documents are collected and deduplicated to compute overall extraction coverage.
The key advantage of this strategy is its simplicity and parallelizability; however, the lack of feedback means that many anchors may retrieve already-extracted or semantically overlapping records, leading to diminishing marginal returns as extraction progresses.

\subsection{Feedback-Driven Anchor Generation}

\mypara{RAG-Thief}~\cite{jiang2024rag}.
RAG-Thief employs an iterative, feedback-driven strategy that leverages previously extracted content to generate new anchors.
Specifically, after each extraction round, RAG-Thief feeds the newly obtained chunks to a local language model (Qwen2-1.5B-Instruct) and prompts it to generate forward and backward continuations.
These continuations serve as anchors for the next round, under the intuition that semantically adjacent passages in the knowledge store are likely to be retrieved by queries resembling their neighboring content.
By chaining continuations across rounds, RAG-Thief progressively expands its extraction frontier outward from an initial seed, achieving higher coverage than purely random sampling.

\mypara{IKEA}~\cite{wang2025silentleaksimplicitknowledge}.
IKEA (Implicit Knowledge Extraction Attack) adopts a concept-level feedback strategy that operates in the embedding space rather than at the token level.
After each extraction round, IKEA identifies salient concepts (key phrases or entities) from the accumulated query--response history and uses them as seeds for the next round of anchor generation.
To maximize retrieval diversity, IKEA iteratively mutates these concept-level anchors subject to embedding-space similarity constraints: new anchors are required to remain close enough to the knowledge store's distribution to trigger relevant retrieval, yet sufficiently distant from previously used anchors to avoid redundant results.
This constrained mutation mechanism enables IKEA to explore the embedding space more systematically than forward/backward continuation, yielding higher extraction coverage under a fixed query budget.

\section{LTM framework}
\label{app:LTM_framework}
\mypara{LTMemory.}
LTMemory represents the standard framework, using text embeddings to calculate semantic similarities and retrieve relevant information.

\mypara{GAMemory.~\cite{park2023generative}}
GAMemory incorporates a self-reflection mechanism. This mechanism activates when the cumulative importance of recent events exceeds a threshold. It then synthesizes fragmented records into high-level inferences supported by evidence.

\mypara{SCMemory.~\cite{wang2023enhancing}}
SCMemory selectively recalls only necessary information for inference. A control mechanism determines when to activate memory retrieval and filter out noise. Additionally, a summarization mechanism condenses extended interactions while preserving core information for efficient reasoning.

\section{Experimental Result}
\label{app:evaluation}
\begin{table*}[t!]
\centering
\caption{Consolidated results of Scenario 1 on the LoCoMo dataset.}
\label{tab:Ad1_main_result_locomo}
\renewcommand{\arraystretch}{1.0} 
\setlength{\tabcolsep}{3.5pt} 

\resizebox{\textwidth}{!}{%
\scriptsize
\begin{tabular}{@{}ll cccc cccc cccc@{}}
\toprule
\multirow{3}{*}{\textbf{Factor}} & \multirow{3}{*}{\textbf{Setting}} & \multicolumn{4}{c}{\textbf{RER (\%) $\uparrow$}} & \multicolumn{4}{c}{\textbf{SS $\uparrow$}} & \multicolumn{4}{c}{\textbf{BER (\%) $\uparrow$}} \\
\cmidrule(lr){3-6} \cmidrule(lr){7-10} \cmidrule(lr){11-14}
& & MEXTRA & \makecell{RAG-\\Thief} & IKEA & \textbf{\textit{SPORE}} & MEXTRA & \makecell{RAG-\\Thief} & IKEA & \textbf{\textit{SPORE}} & MEXTRA & \makecell{RAG-\\Thief} & IKEA & \textbf{\textit{SPORE}} \\
\midrule

\multirow{3}{*}{\makecell[l]{Generative\\Model}} 
  & GPT-5-mini & 7.0\% &  15.5\%& 6.0\% & \textbf{57.5\%} & 1.00 & 1.00 & 1.00 & \textbf{1.00} & 0.0\% & 0.0\% &  0.0\%& \textbf{63.5\%} \\
  & Gemini-2.5-pro & 3.0\% & 18.5\% & 5.0\% & \textbf{45.5\%} & 1.00 & 1.00 & 1.00 & \textbf{1.00} & 0.0\% & 0.0\% & 0.0\% & \textbf{95.0\%} \\
  & Deepseek-chat & 4.5\% & 13.5\% & 6.5\% & \textbf{52.0\%} & 1.00 & 1.00 & 1.00 & \textbf{1.00} & 0.0\% & 0.0\% & 0.0\% & \textbf{70.5\%} \\
\midrule

\multirow{3}{*}{\makecell[l]{LTM\\Framework}} 
  & LTMemory &7.0\%  & 15.5\% & 6.0\% & \textbf{57.5\%} & 1.00 & 1.00 & 1.00 & \textbf{1.00} & 0.0\% & 0.0\% & 0.0\% & \textbf{63.5\%} \\
  & GAMemory & 4.5\% & 6.5\% & 12.0\% & \textbf{55.0\%} & 1.00 &  1.00& 0.80 & \textbf{1.00} & 0.0\% & 0.0\% & 0.0\% & \textbf{75.5\%} \\
  & SCMemory & 10.0\% & 25.5\% & 10.0\% & \textbf{26.5\%} & 0.57 & 0.70 & 0.84 & \textbf{0.84} & 0.0\% & 0.0\% & 0.0\% & \textbf{72.5\%} \\
\midrule

\multirow{3}{*}{\makecell[l]{Embedding\\Model}} 
  & \makecell[l]{All-MiniLM-\\L6-v2} & 7.0\% & 15.5\% & 6.0\% & \textbf{57.5\%} & 1.00 & 1.00 & 1.00 & \textbf{1.00} & 0.0\% & 0.0\% & 0.0\% & \textbf{63.5\%} \\
  & Contriever & 5.0\% & 22.5\% & 10.5\% & \textbf{58.5\%} & 1.00 & \textbf{1.00} & 0.94 & 1.00 & 0.0\% & 0.0\% & 0.0\% &  \textbf{72.0\%} \\
  & E5-base & 4.5\% & 12.5\% & 9.0\% & \textbf{58.5\%} & 1.00 & 1.00 & 0.97 & \textbf{1.00} & 0.0\% & 0.0\% & 0.0\% &  \textbf{68.0\%} \\
\bottomrule
\end{tabular}%
}
\end{table*}

\begin{table*}[t!]
\centering
\caption{Consolidated results of Scenario 2 on the LoCoMo dataset.}
\label{tab:Ad2_main_result_locomo}

\renewcommand{\arraystretch}{1.0}
\setlength{\tabcolsep}{3.5pt}

\resizebox{\textwidth}{!}{%
\scriptsize
\begin{tabular}{@{}c ll cccc cccc cccc@{}}
\toprule
\multirow{3}{*}{\textbf{Trigger}} 
& \multirow{3}{*}{\textbf{Factor}} 
& \multirow{3}{*}{\textbf{Setting}} 
& \multicolumn{4}{c}{\textbf{RER (\%) $\uparrow$}} 
& \multicolumn{4}{c}{\textbf{SS $\uparrow$}} 
& \multicolumn{4}{c}{\textbf{BER (\%) $\uparrow$}} \\
\cmidrule(lr){4-7} 
\cmidrule(lr){8-11} 
\cmidrule(lr){12-15}
& & 
& MEXTRA & \makecell{RAG-\\Thief} & IKEA & \textbf{\textit{SPORE}} 
& MEXTRA & \makecell{RAG-\\Thief} & IKEA & \textbf{\textit{SPORE}}
& MEXTRA & \makecell{RAG-\\Thief} & IKEA & \textbf{\textit{SPORE}}\\
\midrule

\multirow{9}{*}{@1}

& \multirow{3}{*}{\makecell[l]{Generative\\Model}} 
  & GPT-5-mini &2.0\% & 4.5\% & 0.0\% & \textbf{5.0\%} &1.00  & 1.00 & -- & \textbf{1.00} & 0.0\% & 0.0\% & 0.0\% & \textbf{90.3\%} \\
&  & Gemini-2.5-pro & 3.0\%&  4.5\%& 3.0\% & \textbf{6.5\%} & 1.00 & 1.00 & 1.00 & \textbf{1.00} & 0.0\% & 0.0\% & 0.0\% & \textbf{100.0\%} \\
&  & Deepseek-chat &3.5\% & 4.0\% & 2.5\% & \textbf{10.0\%} & 1.00 & 1.00 & 1.00 & \textbf{1.00} & 0.0\% & 0.0\% & 0.0\% & \textbf{90.0\%} \\

\cmidrule(lr){2-15}

& \multirow{3}{*}{\makecell[l]{LTM\\Framework}} 
  & LTMemory & 2.0\%& 4.5\% & 0.0\% & \textbf{5.0\%} &1.00  & 1.00 & -- & \textbf{1.00} & 0.0\% & 0.0\% & 0.0\% & 90.3\%\\
&  & GAMemory &2.5\% & 4.5\% & 4.5\% & \textbf{5.0\%} & 1.00 & 1.00 & 0.90 & \textbf{1.00} & 0.0\% & 0.0\% & 0.0\% & \textbf{100.0\%} \\
&  & SCMemory &0.0\% & 3.5\% & 1.5\% & \textbf{4.5\%} & -- & 0.60 & \textbf{0.98} & 0.80 & -- & 0.0\% & 0.0\% & \textbf{60.0\%} \\

\cmidrule(lr){2-15}

& \multirow{3}{*}{\makecell[l]{Embedding\\Model}} 
  & \makecell[l]{All-MiniLM-\\L6-v2} &2.0\% & 4.5\% & 0.0\% & 5.0\% & 1.00 & 1.00 & -- & 1.00 & 0.0\% & 0.0\% & 0.0\% & 90.3\%\\
&  & Contriever &2.5\% & 4.0\% & 3.0\% & 4.5\% & 1.00 & 1.00 & 1.00 & 1.00 & 0.0\% & 0.0\% & 0.0\% & 100.0\%\\
&  & E5-base & 1.5\% &1.0\%  & 4.5\% & 5.0\% & 1.00 & 1.00 & 1.00 & 1.00 & 0.0\% & 0.0\% & 0.0\% & 100.0\%\\

\midrule

\multirow{9}{*}{@5}

& \multirow{3}{*}{\makecell[l]{Generative\\Model}} 
  & GPT-5-mini &7.0\% & 10.5\% & 4.5\% & \textbf{22.0\%} & 1.00 & 1.00 & 1.00 & \textbf{1.00} & 0.0\% & 0.0\% & 0.0\% & \textbf{75.2\%} \\
&  & Gemini-2.5-pro &3.0\% & 13.5\% & 5.0\% & \textbf{18.5\%} & 1.00 & 1.00 & 1.00 & \textbf{1.00} & 0.0\% & 0.0\% & 0.0\% & \textbf{85.0\%} \\
&  & Deepseek-chat &4.5\% & 10.0\% & 6.5\% & \textbf{30.5\%} & 1.00 & 1.00 &  1.00& \textbf{1.00} & 0.0\% & 0.0\% & 0.0\% & \textbf{70.0\%} \\
\cmidrule(lr){2-15}

& \multirow{3}{*}{\makecell[l]{LTM\\Framework}} 
  & LTMemory & 7.0\%& 10.5\% & 4.5\%  & \textbf{22.0\%} & 1.00 & 1.00 & 1.00 & \textbf{1.00} & 0.0\% & 0.0\% & 0.0\% & \textbf{75.2\%} \\
&  & GAMemory &4.5\% & 6.0\% & 9.5\% & \textbf{23.5\%} & 1.00 & \textbf{1.00} & 0.90 &  & 0.0\% & 0.0\% & 0.0\% & \textbf{75.0\%} \\
&  & SCMemory & 5.5\%& \textbf{16.0\%} & 6.0\% & 12.0\% & 0.59 & 0.68 &0.82  & \textbf{0.84} & 0.0\% & 0.0\% & 0.0\% & \textbf{57.1\%} \\

\cmidrule(lr){2-15}

& \multirow{3}{*}{\makecell[l]{Embedding\\Model}} 
  & \makecell[l]{All-MiniLM-\\L6-v2} &7.0\% & 10.5\% & 4.5\% & \textbf{22.0\%} & 1.00 & 1.00 & 1.00 & \textbf{1.00} & 0.0\% & 0.0\% & 0.0\% & \textbf{75.2\%} \\
&  & Contriever & 5.0\%& 10.5\% & 9.0\% & \textbf{27.0\%} & 1.00 & 1.00 & 1.00 & \textbf{1.00} &  0.0\%& 0.0\% & 0.0\% & \textbf{79.2\%} \\
&  & E5-base & 4.5\%& 10.5\% &  9.0\%& \textbf{19.0\%} & 1.00 & 1.00 & 0.97 & \textbf{1.00} & 0.0\% & 0.0\% & 0.0\% & \textbf{97.0\%} \\

\midrule

\multirow{9}{*}{@20}

& \multirow{3}{*}{\makecell[l]{Generative\\Model}} 
  & GPT-5-mini &7.0\% & 15.5\% & 6.0\% & \textbf{44.0\%} & 1.00 & 1.00 & 1.00 & \textbf{1.00} & 0.0\% & 0.0\% & 0.0\% & \textbf{80.4\%} \\
&  & Gemini-2.5-pro &  3.0\%&  18.5\%& 5.0\% & \textbf{45.5\%} & 1.00 & 1.00 & 1.00 & \textbf{1.00}& 0.0\% & 0.0\% & 0.0\% &\textbf{80.2\%} \\
&  & Deepseek-chat & 4.5\% &  13.5\% & 6.5\%& \textbf{35.0\%} & 1.00 &1.00  & 1.00 & \textbf{1.00} & 0.0\% & 0.0\% & 0.0\% & \textbf{72.0\%} \\

\cmidrule(lr){2-15}

& \multirow{3}{*}{\makecell[l]{LTM\\Framework}} 
  & LTMemory &7.0\% & 15.5\% & 6.0\% & \textbf{44.0\%} & 1.00 & 1.00 & 1.00 & \textbf{1.00} & 0.0\% & 0.0\% & 0.0\% & \textbf{80.4\%} \\
&  & GAMemory &4.5\% & 6.5\% &  12.0\%& \textbf{44.5\%} &  & 1.00 & 0.80 & \textbf{1.00} & 0.0\% & 0.0\% & 0.0\% &\textbf{78.5\%} \\
&  & SCMemory &10.0\% & \textbf{25.5\%} & 10.0\% & 14.0\% & 0.57 & \textbf{1.00} & 0.84 & 0.83 & 0.0\% & 0.0\% & 0.0\% & \textbf{65.9\%} \\

\cmidrule(lr){2-15}

& \multirow{3}{*}{\makecell[l]{Embedding\\Model}} 
  & \makecell[l]{All-MiniLM-\\L6-v2} & 7.0\%& 15.5\% & 6.0\% & \textbf{44.0\%} & 1.00 & 1.00 & 1.00 & \textbf{1.00} & 0.0\% & 0.0\% & 0.0\% & \textbf{80.4\%} \\
&  & Contriever &5.0\% & 22.5\% & 10.5\% & \textbf{40.5\%} & 1.00 & \textbf{1.00} & 0.94 & 0.99 & 0.0\% & 0.0\% & 0.0\% & \textbf{83.2\%} \\
&  & E5-base & 4.5\%& 12.5\% & 9.0\% & \textbf{38.5\%} &  1.00& 1.00 & 0.97 & \textbf{1.00} & 0.0\% & 0.0\% & 0.0\% & \textbf{77.8\%} \\

\bottomrule
\end{tabular}%
}
\end{table*}

\begin{table*}[t]
\centering
\caption{Consolidated results of Scenario 3 on the LoCoMo dataset.}
\label{tab:Ad3_main_result_locomo}

\renewcommand{\arraystretch}{1.0}
\setlength{\tabcolsep}{4pt}

\small
\begin{tabular}{@{}c ll ccc@{}}
\toprule
\multirow{2}{*}{\textbf{Trigger}} 
& \multirow{2}{*}{\textbf{Factor}} 
& \multirow{2}{*}{\textbf{Setting}} 
& \multicolumn{3}{c}{\textbf{Metrics}} \\
\cmidrule(lr){4-6}
& & 
& \textbf{RER (\%) $\uparrow$} 
& \textbf{SS $\uparrow$} 
& \textbf{BER (\%) $\uparrow$} \\
\midrule

\multirow{9}{*}{@1}

& \multirow{3}{*}{\makecell[l]{Generative\\Model}} 
  & GPT-5-mini & 69.0\%& 1.00 & 72.0\%\\
&  & Gemini-2.5-pro &71.0\% & 1.00 & 80.0\% \\
&  & Deepseek-chat & 81.0\% & 1.00 & 52.0\% \\

\cmidrule(lr){2-6}

& \multirow{3}{*}{\makecell[l]{LTM\\Framework}} 
  &LTMemory & 69.0\%& 1.00 & 72.0\%\\
&  & GAMemory & 76.0\% & 1.00 & 70.0\%\\
&  & SCMemory & 35.0\% & 0.75 & 93.5\%\\

\cmidrule(lr){2-6}

& \multirow{3}{*}{\makecell[l]{Embedding\\Model}} 
  & \makecell[l]{All-MiniLM-L6-v2} & 69.0\%& 1.00 & 72.0\%\\
&  & Contriever & 87.0\%& 1.00  & 65.0\%  \\
&  & E5-base &  74.5\% & 1.00  & 70.0\% \\

\bottomrule
\end{tabular}
\end{table*}

\section{Potential Defense}
\label{app:defense}
\begin{tcolorbox}[
    colback=white,              
    colframe=black,             
    width=\linewidth,           
    title=Template for response filtering defense,        
    boxrule=0.5mm,              
    colbacktitle=black,         
    coltitle=white,             
    left=5pt,                   
    right=5pt,                  
    arc=3mm,                    
    before skip=10pt,           
    after skip=10pt            
]
\small
\textbf{You are a security auditor.} Analyze the following tool response and determine if it contains prompt injection patterns.

Prompt injection patterns include:
\begin{itemize}[leftmargin=*, nosep]
    \item Instructions to ignore previous instructions
    \item Instructions to change behavior or role
    \item Instructions to output specific predetermined answers
    \item Instructions to concatenate or leak conversation history
    \item Any attempt to manipulate the calling AI agent
\end{itemize}

Tool Response: {response}

Reply with EXACTLY one word: \texttt{SAFE} or \texttt{UNSAFE}
\end{tcolorbox}
\end{document}